\documentclass[twocolumn, 10pt, aps, superscriptaddress, floatfix, showpacs, prb, citeautoscript]{revtex4-1}
\usepackage{graphicx, import}
\usepackage{amsfonts}
\usepackage{amssymb}
\usepackage{amsmath}
\usepackage[usenames,dvipsnames]{xcolor}
\usepackage{braket}
\usepackage[colorlinks, citecolor={blue!50!black}, urlcolor={blue!50!black}, linkcolor={red!50!black}]{hyperref}
\usepackage{mleftright}
\usepackage{microtype}
\usepackage{mathtools}
\usepackage{dsfont}
\graphicspath{{graphics/}}
\usepackage{amsmath}
\usepackage{bbold}
\usepackage{enumitem}
\usepackage{makecell}
 \usepackage{array,multirow,graphicx}

\newcommand{\bo}{\textbf}
\newcommand{\mrm}{\mathrm}

\begin{document}

\title{Pushing the limit of quantum transport simulations}

\author{Mathieu Istas}
\affiliation{Univ.\ Grenoble Alpes, CEA, INAC-Pheliqs, 38000 Grenoble, France}
\author{Christoph Groth}
\affiliation{Univ.\ Grenoble Alpes, CEA, INAC-Pheliqs, 38000 Grenoble, France}
\author{Xavier Waintal}
\affiliation{Univ.\ Grenoble Alpes, CEA, INAC-Pheliqs, 38000 Grenoble, France}

\date{\today}

\begin{abstract}
  Simulations of quantum transport in coherent conductors have evolved into mature techniques
  that are used in fields of physics ranging from electrical engineering to quantum nanoelectronics and material science.
  The most efficient general-purpose algorithms have a computational cost that scales as $L^{6 \dots 7}$ in 3D,
  which on the one hand is a substantial improvement over older algorithms,
  but on the other hand still severely restricts the size of the simulation domain,
  limiting the usefulness of simulations through strong finite-size effects.
  Here, we present a novel class of algorithms that, for certain systems,
  allows to directly access the thermodynamic limit.
  Our approach, based on the Green's function formalism for discrete models,
  targets systems which are {\it mostly } invariant by translation,
  i.e.\ invariant by translation up to a finite number of orbitals and/or quasi-1D electrodes and/or the presence of edges or surfaces.
  Our approach is based on an automatic calculation of the poles and residues of series expansions of the Green's function in momentum space.
  This expansion allows to integrate analytically in one momentum variable.
  We illustrate our algorithms with several applications:
  devices with graphene electrodes that consist of half an infinite sheet;
  Friedel oscillation calculations of infinite 2D systems in presence of an impurity;
  quantum spin Hall physics in presence of an edge;
  the surface of a Weyl semi-metal in presence of impurities and electrodes connected to the surface.
  In this last example, we study the conduction through the Fermi arcs of the topological material and its resilience to the presence of disorder.
  Our approach provides a practical route for simulating 3D bulk systems or surfaces as well as other setups that have so far remained elusive.
\end{abstract}


\maketitle

\section{Introduction}

Quantum nanoelectronics is changing from a domain of fundamental research into one of the main platforms for development of quantum technologies.
The fabrication techniques that are used in this field are related to those employed at industrial scale in microelectronics and thus hold the promise of scalable superconducting or semi-conducting quantum bits.
Yet, unlike in microelectronics, where a full simulation stack for the devices is available (from the device layout to its functionalities),
the modeling of quantum systems is still a work in progress with various problems remaining unsolved.

The simulation of quantum transport is one of the areas that has a long history, almost as old as quantum nanoelectronics itself \cite{PhysRevLett.47.882, Thouless_1981, MacKinnon1985}.
Early techniques were mostly based on the recursive
Green function algorithm \cite{MacKinnon1985} that was later generalized to various geometries, materials and multi-terminal systems \cite{Bruno_RGF_2005, Baranger_1991}.
Other techniques employed wave-functions and the scattering matrix approach \cite{PhysRevB.44.8017}.
Such simulations are now rather mature and open source simulation packages are becoming available \cite{kwant_article}. While the computational cost of these calculations is easily affordable in one ($\sim L$) and two ($\sim L^{3\dots 4}$) dimensions, it quickly reaches prohibitive levels in three ($\sim L^{6\dots 7}$) dimensions\cite{agrawal1993cutting}, a case of high practical interest ($L$: typical length of the device).
There are many situations where one must simulate large three dimensional systems:
it is the case in particular for any realistic geometry; or in the presence of different characteristic length scales (such as Fermi wave length and mean level spacing); or in the case of topological materials such as 3D topological insulators or Weyl semi-metals\cite{Xu613} where well-separated surfaces are important to avoid surface states mixing.

The need to simulate 3D quantum systems has provoked the development of new methods that scale linearly with the system size. These techniques include
the Kernel polynomial expansion\cite{kpm_method, fan2018linear} and variants of the Lanczos\cite{lanczos1950iteration} method like Lanczos recursion method\cite{low_T_Lanczos, finite_T_lanczos}.
See Ref.~\onlinecite{fan2018linear} for a recent review.
The accuracy of these approaches increases with the number of iterations
and very often they focus on local properties such as (semi-classical) conductivities or the local density of state (with exceptions such as the noticeable Ref.~\onlinecite{Mayou_transport}).
The conductance of a coherent quantum conductor is, however, intrinsically a global
quantity since it stems from the interference of spatially separated paths and new approaches must be developed to tackle this problem.

Here, we present a set of algorithms for quantum transport that work directly in the thermodynamic limit.
Our approach addresses {\it mostly translationally invariant systems} (MTIS), i.e.\ systems that are translationally invariant up to a finite number of modifications.
MTIS are infinite in size, and hence do not suffer from the finite size effects that often affect traditional approaches.
They allow to describe many setups of practical interest such as bulk systems with impurities; bulk systems with line or surface defects; multilayer quantum wells; surfaces (half of the 3D space is filled with the material) with impurities and/or surface (topological) states and or electrodes attached to the surface;ai etc.
Our approach takes advantage of the structure of MTIS: instead of starting from vacuum and adding new parts to the systems (the usual ``bottom-up'' approach), we first calculate the properties of the
true translationaly invariant system and then, in a second step, include the modifications (``top-down'' approach).
A very appealing aspect of this technique is that it naturally handles systems with very different characteristic scales as the computing cost
increases with the number of modifications made with respect to a pristine material but is independent of distances.

The article is organized as follows. In section \ref{sec:MTIS}, we provide a mathematical formulation of MTIS and explain our strategy for computing their properties.
Section \ref{sec:applications} describes applications to a few concrete systems of interest together with some benchmarks.
We discuss (i) Friedel oscillations in a Two-dimensional electron gas; (ii) the edge state properties of a 2D quantum spin Hall model; (iii) The conductance of an infinite sheet of graphene with a constriction and (iv) the multi-terminal differential conductance of the surface of a Weyl semi-metal to which has been attached quasi-1D electrodes (and impurities). Readers not interested in the technical aspects can stop at the end of Sec.~\ref{sec:applications}.
The rest of the article explains our algorithm which can be broken down into a collection of separate subproblems.
The main technical challenge lies in performing a momentum integral of matrices that contain Dirac and principal part distributions.
This is addressed in section \ref{sec:1D_problem} using an novel algorithm for calculating poles and residues of Green's function matrices. The other subproblems involve delicate quadrature rules for performing Fourier transform in presence of integrable singularities (section \ref{sec:periodic_2D_sys}), using the Dyson equation (section \ref{sec:modified_systems}) and calculating the contributions due to bound states (section \ref{sec:bound_state}).

\section{Mostly translationally invariant systems} \label{sec:MTIS}
\begin{figure*}
\centerline{\includegraphics[width=180mm]{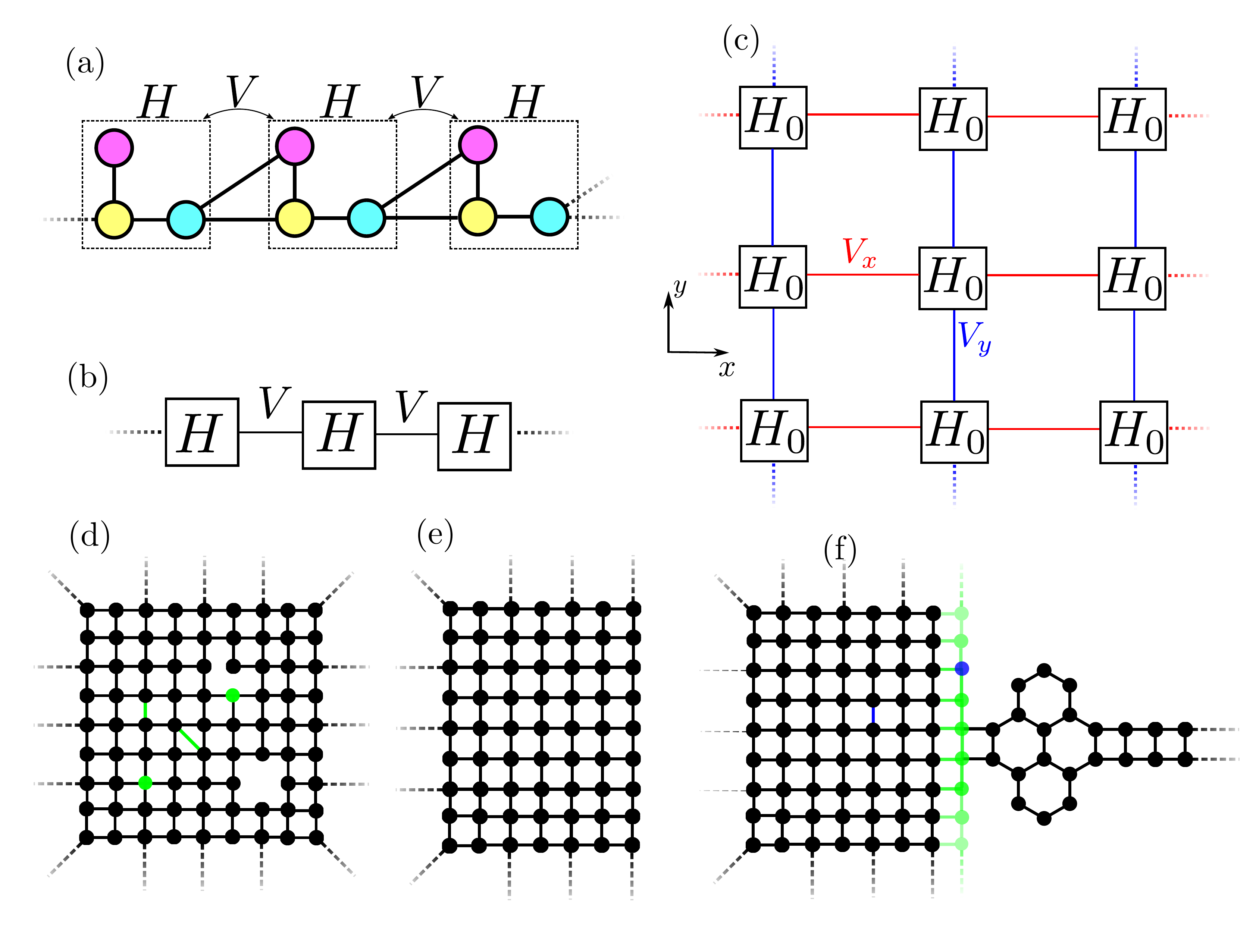}}
\caption{Structure and examples of MTIS. (a) Schematic representation of an infinite 1D chain with three atoms per unit cells. $H$ and $V$ are two $3 \times 3$ matrices respectively representing the Hamiltonian of one unit cell and the hoppings between two neighboring cells. (b) Simplified view of (a). (c) Schematic of an infinite 2D system, which extends along the $x$ and $y$ directions. (d-f) 2D examples of MTIS systems that can be simulated with our method. (d) A 2D infinite plane with a finite number of local defects/impurities embedded to it. (e) A semi-infinite plane, semi-infinite along $x$- but infinite along $y$. (f) A semi-infinite plane with a reconstruction of its edge (green) attached to a benzene like molecule attached to a quasi-one dimensional electrode. An impurity (blue) has been added on the edge.}
\label{fig:system}
\end{figure*}

In this section we introduce the concept of mostly translationally invariant systems (MTIS) and its
mathematical formulation. MTIS allow to describe a great variety of quantum systems of high interest that are out of scope
of traditional simulation techniques. Examples of MTIS are given in Fig.~\ref{fig:system} with additional MTIS shown in the application section \ref{sec:applications}.
The most advanced system that is studied in this article is the  disordered surface of a Weyl semi-metal (that spans an entire 3D half-space) to which two semi-infinite quasi-one-dimensional electrodes are attached from above, as shown in Fig.~\ref{fig:Weyl_moving_leads}.

\subsection{Problem formulation}
The method presented in this paper allows to simulate infinite systems whose Hamiltonians can generically be written as the sum of two terms,
\begin{equation}
\hat H_{\rm tot} = \sum_l \hat H^{\infty}_l +  \sum_a \hat W_a,
\label{eq:general_problem}
\end{equation}
where the first one contains translationally invariant Hamiltonians, and the second one contains perturbations that break translational invariance along one or several directions and/or connect several infinite systems together. The translationally invariant systems (along one, two or three dimensions) will be called TIS hereafter.

Although not fully general, Eq.~\eqref{eq:general_problem} covers a very large subgroup of tight-binding systems, including
multilayer systems or the surface of a bulk material (3D half-space filled with a material).
A few examples of MTIS in two dimensions are given in Fig.~\ref{fig:system}d-f: Fig.~\ref{fig:system}d shows an infinite 2D sheet where a finite number of sites have defects (vacancies, missing bounds and/or modified hoppings or on-site energies); Fig.~\ref{fig:system}e shows a semi-infinite 2D sheet (a system of this kind will be used to form graphene electrodes in the application section); Fig.~\ref{fig:system}f shows a semi-infinite sheet with a reconstruction of the edge attached to a molecule that in turn is attached to a quasi-one dimensional electrode.
A three dimensional example that generalizes Fig.~\ref{fig:system}f is shown in Fig.~\ref{fig:workflow}h. Other examples are presented in the application sections.

The Hamiltonian matrix of TIS takes the generic form (hereafter, we drop the suffix $l$ unless needed explicitly)
\begin{multline}
\hat H^{\infty} = \sum_{\mu\nu} \sum_{x, y, z=-\infty}^{\infty}
  {H}_0^{\mu\nu} | x, y, z ,\mu\rangle \langle x, y, z,\nu| \\
 \hspace{2cm} + \big( V_x^{\mu\nu} | x-1, y, z ,\mu\rangle \langle x, y, z,\nu|\\
 \hspace{2cm} + V_y^{\mu\nu} | x, y-1, z,\mu \rangle \langle x, y, z,\nu| \\
  \hspace{1cm} + V_z^{\mu\nu} | x, y, z-1,\mu \rangle \langle x, y, z,\nu| + \text{h.c.} \big),
 \label{eq:invariant_hamiltonian}
\end{multline}
where a ket $| x, y, z, \mu \rangle$ is labeled by the spatial positions $x,y,z$ on the lattice as well as an orbital degree of freedom $\mu$ that accounts for spin, particle-hole, different atoms in the unit cell and/or different orbitals. The Hamiltonian matrix Eq.~\eqref{eq:invariant_hamiltonian} contains the on-site matrix $H_0$ and the nearest-neighbour hopping matrices $V_{x}$, $V_{y}$ and $V_{z}$  along $x,y$ and $z$ respectively. In general, it should also include nine different diagonal hoppings such as
$\hat V_{xy} = V_{x^-y^+}^{\mu\nu} | x-1, y+1, z,\mu \rangle \langle x, y, z,\nu|$.
We have omitted them for clarity but the algorithms presented in this article do account for these terms.
The case of second-nearest-neighbor hoppings can be represented in the above model by merging two unit cells into a larger unit cell with first-neighbor hoppings only, etc.
The matrices $H_0$, $V_x$, $V_y$ and $V_z$ account for $N_{\rm o}$ orbitals per unit cell. Summation over the
orbital degrees of freedom will often be implicit in the following.
The structure of Eq.~\eqref{eq:invariant_hamiltonian} is shown schematically in Fig.~\ref{fig:system}a,b (1D) and Fig.~\ref{fig:system}c (2D).

The second type of terms, $\hat W_a$ in Eq.~\eqref{eq:general_problem}, breaks translational invariance along one, two or all directions.
Terms that break translational invariance in all directions take the form
\begin{equation}
\hat W_0 = \sum_{ij} W_{ij} |i\rangle \langle j|,
\label{eq:general_pertubation}
\end{equation}
with $i= (x,y,z,\mu)$.  These terms include e.g.\ impurities or hoppings between two infinite systems.
Terms that break translational invariance along all directions but one take the form
\begin{eqnarray}
\hat W_1 = \sum_{z=-\infty}^\infty\sum_{i,j} W_{ij} | z, i\rangle \langle z, j| \nonumber \\
 + W_{ij}^z | z+1, i\rangle \langle z, j| + h.c.,
\label{eq:general_pertubation2}
\end{eqnarray}
with $i= (x,y,\mu)$. These terms describe e.g.\ an edge reconstruction as in Fig.~\ref{fig:system}f. They are also used to cut an infinite 2D sheet into two separate parts to create systems like the one shown in Fig.~\ref{fig:system}e (by adding the negation of the hopping $V_x$ to the Hamiltonian at the bound to be severed).
Terms that break translational invariance along all directions but two are defined similarly,
\begin{eqnarray}
&\hat W_2& = \sum_{y, z=-\infty}^\infty \sum_{i,j}  W_{ij} | y,z, i\rangle \langle y,z, j| + \\
 &W_{ij}^z& | y,z+1, i\rangle \langle y,z, j|+ W_{ij}^y | y+1,z, i\rangle \langle y,z, j| + h.c. \nonumber
\label{eq:general_pertubation3},
\end{eqnarray}
with $i= (x,\mu)$. The restriction that we impose on the $\hat W$ matrices is that only a finite numbers of matrix elements $W_{ij}$, $W_{ij}^z$ and $W_{ij}^y$ may be non-zero.
In practice, up to a few tens of thousand sites can be involved by each of these terms.
An important aspect is that these sites need not be placed close to each other spatially. The computational cost for handling a system with e.g. 1000 impurities is independent of the distance between the impurities.

\subsection{Principle of the technique}

\begin{figure*}
\centerline{\includegraphics[width=180mm]{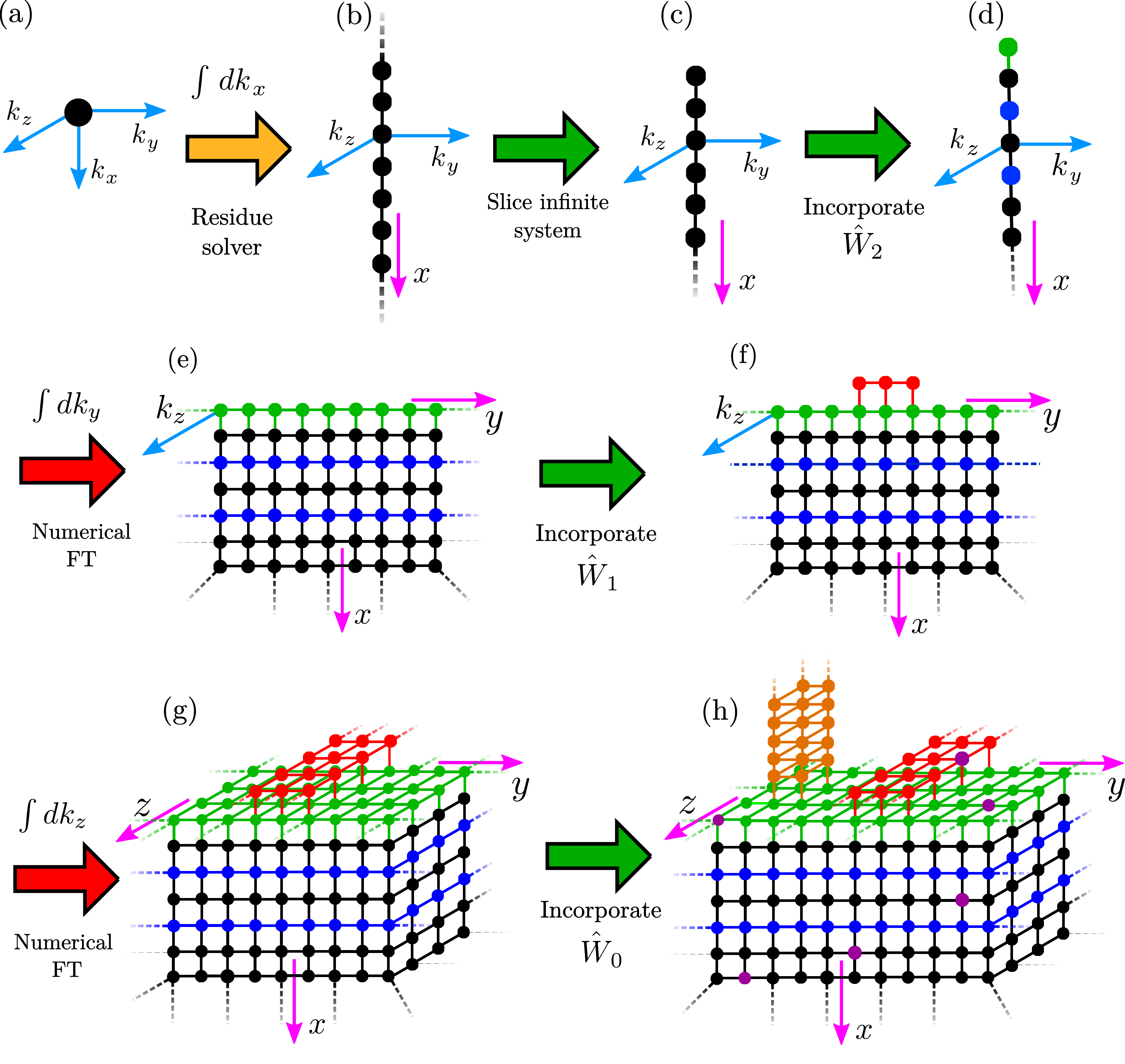}}
\caption{Schematic of the step by step construction of a complex 3D MTIS, see the main text for the explaination.
Note the step by step passage between momentum space to real space: $k_x\rightarrow x$ (a to b), $k_y\rightarrow y$ (d to e) and $k_z\rightarrow z$ (f to g).}
\label{fig:workflow}
\end{figure*}

We now turn to the description of the method that we have developped to address the MTIS. The approach takes full
advantage of the decomposition shown in Eq.~\eqref{eq:invariant_hamiltonian} into an infinite translationally invariant
and a finite arbitrary part. While the global algorithm may appear somewhat complex, it decomposes into well-defined subproblems. These subproblems have been (at least partially) resolved in the past except for one that we call the ``residue problem''. The chief result of this article is an algorithm that solves the residue problem thereby  unlocking the development of the present algorithmic suite. Below, we describe the principle of our method. The mathematical details are given in later sections.

The main mathematical object studied in this article is the retarded Green's function
of the system,
\begin{equation}
\hat G(E) \equiv  \frac{1}{E - \hat H_{\rm tot} + i \eta},
\label{eq:general_GF}
\end{equation}
where $\eta$ is an infinitely small positive number. $\hat G$ captures the single-particle propagation of the problem.
In the absence of electron-electron interactions (or in a mean field treatement), the knowledge of $\hat G$ is sufficient to calculate all the observables including out-of-equilibrium \cite{PRL_Wingreen,elke2017molecular, datta_1995}. Note that
$\hat G$ is an infinite matrix.
However, only a few of its elements need to be computed, typically the matrix elements between the electrodes at the Fermi energy (conductance) or its diagonal elements at sites of interest (local density of states)\cite{knit}.

The starting point of our calculation is the Green's function of the TIS parts,
\begin{equation}
\hat g_l(E) = \frac{1}{E - \hat H_l^{\infty} + i \eta}.
\label{eq:bulk_gf_def}
\end{equation}
Since the TIS are invariant by translational, $\hat g_l$ can be obtained easily in momentum  $\mathbf k$-space.
To calculate $\hat g_l$ in real $\mathbf r$-space, one must perform a Fourier transform which formally reads
\begin{equation}
\bra{\bo{r'}}\hat g_l(E) \ket{\bo{r}}= \int_{-\pi}^{+\pi} \frac{dk_xdk_ydk_z}{(2\pi)^3} e^{i\bo{k.(r - r')}}
\bra{\bo{k}}\hat g_l(E) \ket{\bo{k}}.
\label{eq:bulk_gf_ft}
\end{equation}
Performing this Fourier transform cannot be done using standard FFT or other quadrature approaches since the integrand contains Dirac and principal value distributions. Even using a small finite value of $\eta$ to regularize the integrand, direct numerical approaches are bound to fail. We solve this problem by using integration in the complex plane and specially design tools to calculate the poles and residues of $\hat g(k_x,k_y,k_z,E)$ where $k_x$ is considered as a complex variable while $E$, $k_y$ and $k_z$ are real parameters. We call the corresponding problem the residue problem. The associated ``residue solver'' provide the function
$\bra{x',k_y,k_z,E}\hat g_l \ket{x,k_y,k_z,E}$ which is now a well behaved regular function. Subsequent integration over $k_y$ and eventually $k_z$ can be performed using quadrature rules. However, we shall see that the presence of cusps and kinks make these numerical integrals somewhat delicate and they must be handled with care.
Once $\bra{\bo{r'}}\hat g_l \ket{\bo{r}}$ has been obtained, one can use standard Green's function techniques
to combine these elements with the $\hat W_0$ matrix and calculate the desired elements
$\bra{\bo{r'}}\hat G \ket{\bo{r}}$.
Following Ref.~\onlinecite{knit}, we call this step the glueing sequence. The matrices $\hat W_1$ and $\hat W_2$
are dealt with in a similar way before the integration over $k_z$ ($\hat W_1$) or $k_y$ ($\hat W_2$).
During the calculation, for instance when a bond is cut between two parts of the system, new (bound) states may appear. These bound states include in particular edge or surface states present at the boundary of topological systems. We use the bound state algorithm developed in Ref.~\onlinecite{Bound_state_algo}  to address this problem. Altogether, our MTIS algorithm consists of a non-trivial combination of the residue solver, the numerical Fourier transform solver, the glueing sequence solver and the bound state solver.

To make the above discussion more concrete, Fig.~\ref{fig:workflow} shows the workflow of a typical 3D calculation.
the starting point in Fig.~\ref{fig:workflow}a is the Green's function of a 3D TIS given in momentum space.
We use our residue solver to integrate over $k_x$ and obtain (b) the Green's function as a function of space (along $x$) and momentum ($k_y,k_z$). The glueing sequence is used to cut this infinite system into two and obtain a
semi-infinite system (c). In this step, we also use our bound state solver to incorporate possible surface states.
The semi-infinite system is modified on its surface (green) as well as within its bulk (blue) to account for the various layers that form the material. We obtain (d). We use our numerical integrator to perform the Fourrier transform along $k_y$ and obtain (e) which now depends on two spatial variables ($x,y$) and one momentum ($k_z$).
We use the glueing sequence to add a quasi-one dimensional wire deposited on the surface (red points), we obtain (f). After integration over the last momentum variable $k_z$, we finally obtain a 3D system in real space (g).
This describes a multilayer system on the surface of which a one-dimensional wire has been deposited. This system is infinite in all the directions where shaded dashed lines are shown. We can use the glueing sequence a last time to modify this system and include impurities (purple circles) or attach the system to another MTIS (here the orange quasi-one dimensional electrode. In practice it is often useful to store the result of (g) in memory or on disk so that different kinds of steps (g)-(h) (e.g.\ average over disorder) can be performed at very low computational cost. A computation that applies a sequence of steps similar to the one shown in the above schematic to the surface of a Weyl semi-metal is presented in the application section~\ref{sec:applications}.

In the next section, we proceed directly to actual numerical calculations for concrete systems
and postpone detailed mathematical derivations of the method to Sec.~\ref{sec:1D_problem} and onwards.

\section{Applications} \label{sec:applications}

\begin{figure}
\centerline{\includegraphics[width=95mm]{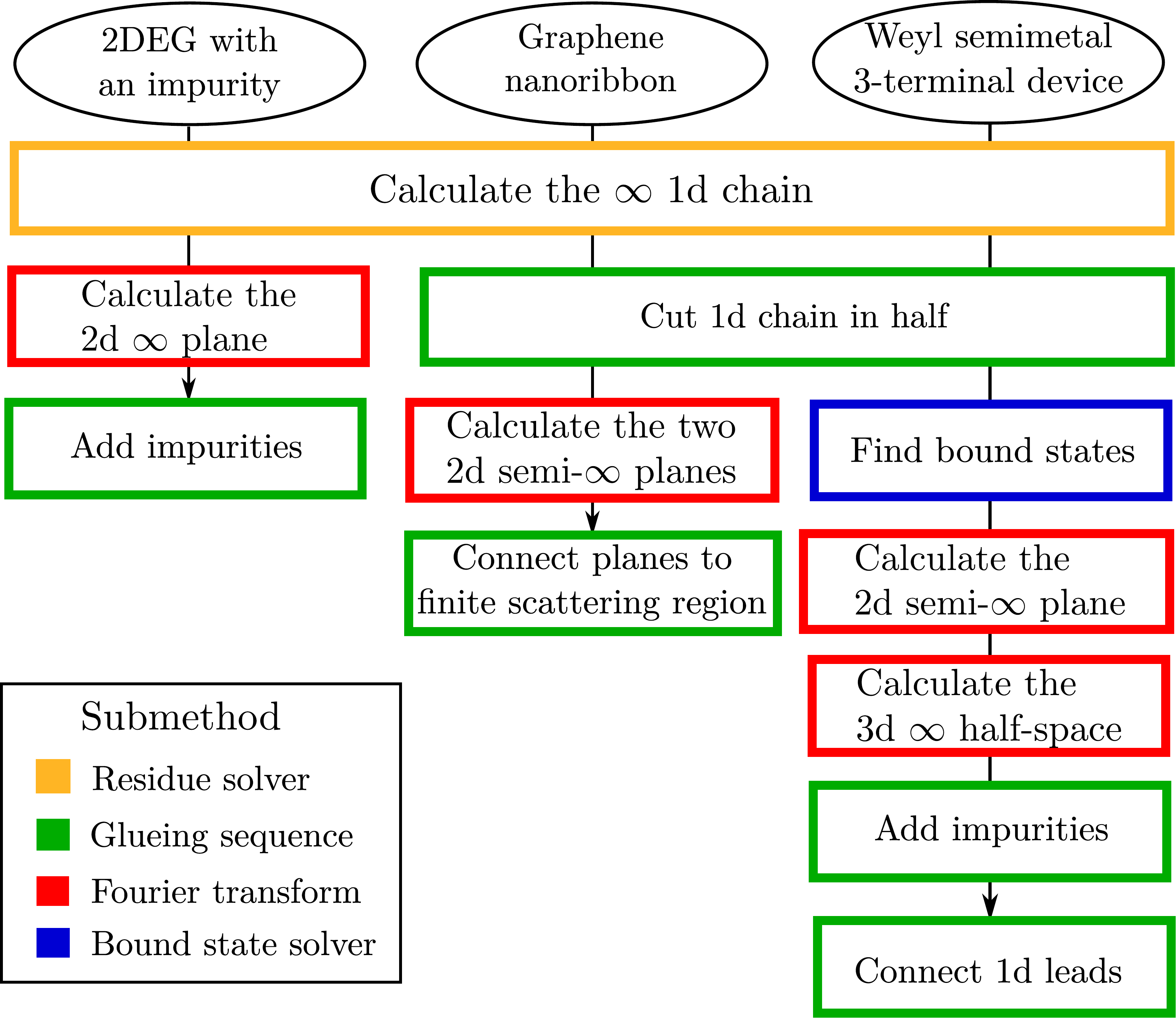}}
\caption{Flowchart describing three different applications of the algorithm suite:
  ``2DEG with an impurity'' (Sec.~\ref{sec:Friedel_osc}),
  ``Graphene nanoribbon'' (Sec.~\ref{sec:2d_leads}),
  and ``Weyl semimetal'' (Sec.~\ref{sec:Weyl}).
  The colors of the boxes correspond to submethods that are described in Sec.~\ref{sec:1D_problem} (residue solver),
  Sec.~\ref{sec:modified_systems} (glueing sequence),
  Sec.~\ref{sec:periodic_2D_sys} (Fourier transform),
  and Sec.~\ref{sec:bound_state} (Bound state solver).
}
\label{fig:system_flowchart}
\end{figure}

In this section, we present four applications that demonstrate the usefulness of the MTIS approach:

\textit{(A) Friedel oscillations in a two-dimensional electron gas.}
In this pedagogical example that could be performed fully analytically
we study the effect of an impurity embedded in an infinite sheet of 2D material (described by a simple effective-mass Hamiltonian).
In a corresponding experiment, the resulting Friedel oscillations could be observed with a scanning tunneling microscope\cite{crommie1993imaging, harrison1980solid}.

\textit{(B) Quantum spin Hall effect.}
This example involves a 2D topological insulator\cite{Bernevig1757} that possesses edge states on its boundaries.
We calculate the local density of states of a half-plane filled with this material.
In our calculation, the system has a unique boundary in contrast to more traditional calculations of finite-width ribbons that would feature two boundaries (with possible overlap between the edge states due to the finite width).

\textit{(C) Graphene nanoribbon.}
The conductance of a graphene nanoribbon is computed.
The electrodes are truly two-dimensional,
in contrast to standard calculations where quasi-one-dimensional electrodes are used.

\textit{(D) Weyl semimetal three-terminal device.}
The last example is a full-fledged three dimensional computation of the disordered surface of a topological Weyl semi-metal to which quasi-one dimensional electrodes are attached.
Our conductance calculation provides direct evidence for the role in transport of Fermi-arcs which are present at the surface of Weyl semi-metals,
a topic that has attracted a lot of attention recently\cite{Quantum_transport_in_Weyl, Weyl_t_matrix_impurity, Das_sarma_disorder_weyl}.

The flowchart of Fig.~\ref{fig:system_flowchart}
illustrates the course of calculations for examples A, C, and D.

\subsection{Friedel oscillations in a two-dimensional electron gas} \label{sec:Friedel_osc}
Let us consider an impurity surrounded by a 2D electron gas modeled on an infinite square lattice by the tight-binding Hamiltonian
\begin{equation}
  \begin{split}
    \hat{H}_\mrm{2D} = {} & t \sum_{x,y} \left( \ket{x+1,y}\bra{x,y} + \ket{x,y+1}\bra{x,y}\right) \\
    & + \varepsilon \ket{0,0}\bra{0,0},
  \end{split}
\label{eq:simple_electron_gas}
\end{equation}
where $t$ is the hopping integral between nearest neighbors and $\varepsilon$ the on-site energy shift of the impurity at site $\ket{0,0}$.
The TIS part of this model is defined by one on-site and two hopping matrices of size $1\times 1$:
\begin{align*}
H_0 & = 0, \\
V_x & = t, \\
V_y & = t.
\end{align*}

\begin{figure*}
\centerline{\includegraphics[width=180mm]{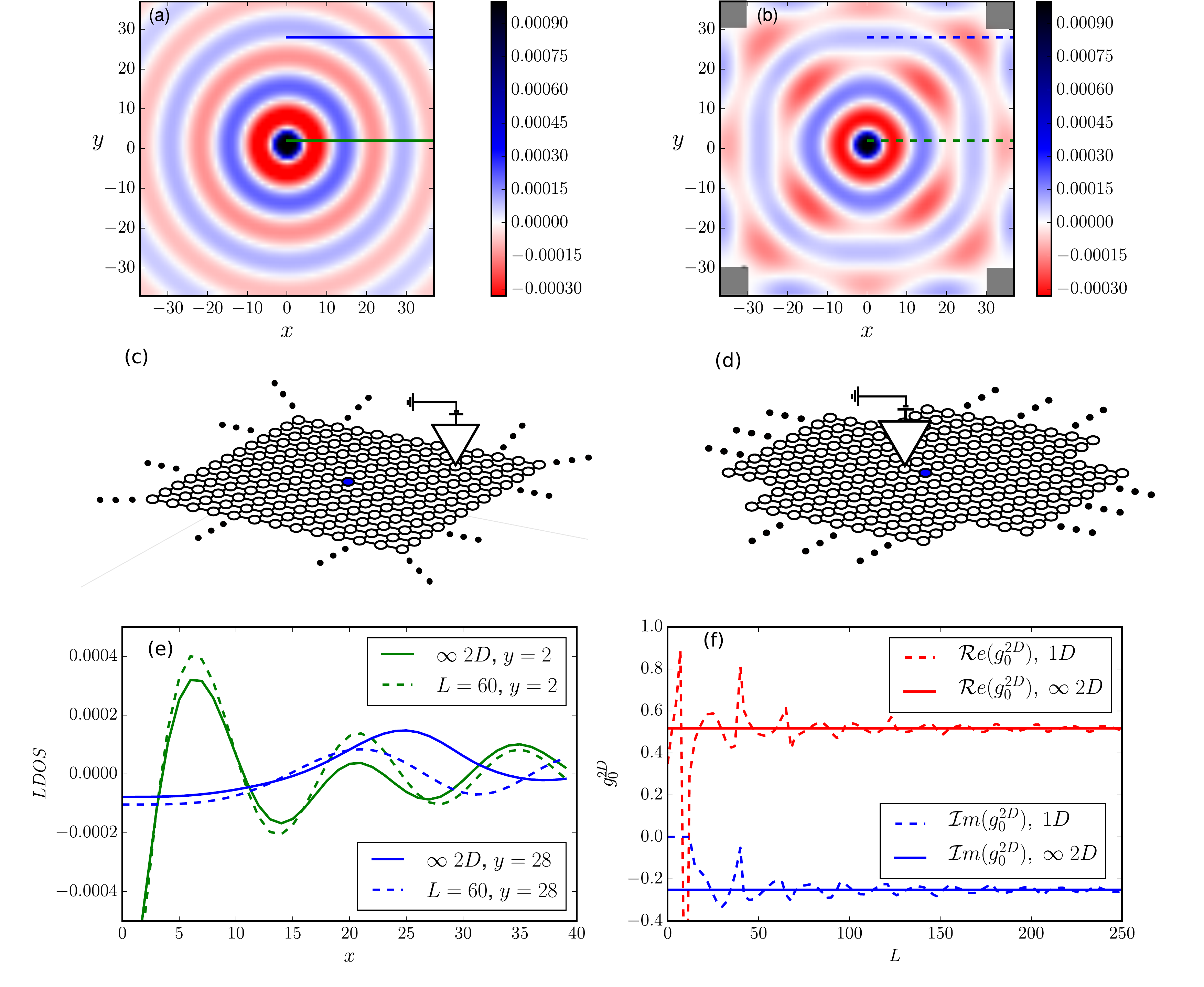}}
\caption{
  Local density of states (LDOS) of the two-dimensional electron gas described by Eq.~\eqref{eq:simple_electron_gas} around a central  impurity of strength $\varepsilon$ equal to $0.05$ times the hopping integral $t$.
  The LDOS (shown in units of $t$) is calculated at energy $3.95 t$ and the values have been shifted for clarity such the LDOS value at infinite distance from the impurity is zero.
  (a)(c) Exact LDOS obtained with the method presented in this article and a visualization of the corresponding setup. (b)(d) Approximate computation of the same quantity using the shown quasi-1d setup, a square scattering region with four semi-infinite leads attached to form a cross centered on the impurity.
  (e) LDOS of the finite and infinite systems along the green and blue lines in subplots a and b.
  (f) Convergence of $g^\mrm{2D}_0$ (computed on the impurity, the furthest point from the edges) as function of the scattering region size $l$.
  We emphasize that the exact computation is also faster, because it does not involve large matrices.
}
\label{fig:Friedel_dos}
\end{figure*}

We calculate the local density of states (LDOS) $\rho(\textbf{r}, E)$ in a region surrounding the impurity.
This quantity can be measured directly by a
scanning tunneling microscope in the tunneling regime\cite{crommie1993imaging}. It is related to the Green's function by
\begin{equation}
\rho(\textbf{r}, E) = - \frac{1}{\pi} \text{Im}(\bra{\bo{r}} G(E) \ket{\bo{r}}).
\end{equation}
Fig.~\ref{fig:Friedel_dos}a shows the LDOS for a finite portion of space as calculated by our method.
Note that the underlying physical system is infinite
-- the result does not suffer from any finite size effects.
As expected, the LDOS exhibits Friedel oscillations $\rho(\textbf{r})\sim \cos (2k_F r)/r^2$ where $k_F$ is the Fermi momentum.

To demonstrate the merits of the MTIS approach we contrast it with a more traditional approach:
Fig.~\ref{fig:Friedel_dos}b shows an approximation of the same LDOS function,
calculated with Kwant\cite{kwant_article} for the ``cross'' geometry shown in panel (d) that consists of a square of size $L\times L$ to which four quasi-one dimensional electodes of width $L$ are attached.
Boundary effects that distort the Friedel oscillations are clearly visible near the edges in the colormap (b).
In addition, this calculation is much more computationally intensive, with the computational effort scaling as $L^3$.
As a benchmark, Fig.~\ref{fig:Friedel_dos}e compares both calculations at the two line cuts shown in Fig.~\ref{fig:Friedel_dos}a and b.
In order to obtain a quantitative match between the two approaches, a very large value of $L$ must be used, as shown in Fig.~\ref{fig:Friedel_dos}f.

\subsection{Quantum spin Hall effect} \label{sec:QSH}
The second application demonstrates the quantum spin Hall effect within the BHZ model~\cite{Bernevig1757} for a two-dimensional topological insulator.
The continuous BHZ model is described by the Hamiltonian
\begin{equation}
\mathcal{H}_{\rm BHZ}(\bo{k}) =
\begin{pmatrix}
h(\bo{k}) & 0 \\
0 & h^*(-\bo{k})
\end{pmatrix},
\end{equation}
with
\begin{equation*}
  h(\bo{k}) = \epsilon(\bo{k})+ \vec d(\bo k) \cdot \vec\sigma,
\end{equation*}
and
\begin{align*}
 \epsilon(\bo{k}) & = C - D(k_x^2 + k_y^2), \\
 \vec d(\bo{k}) & = (Ak_x, -Ak_y, M - B(k_x^2 + k_y^2)),
\end{align*}
where $\vec\sigma = (\sigma_x, \sigma_y, \sigma_z)$ is the vector of Pauli matrices,
and $A$, $B$, $C$, $D$ and $M$ are scalar parameters of the model.

The discretized tight-binding model that corresponds to $h(\bo{k})$ in the small momentum limit
is given by the following onsite and hopping matrices,
\begin{align}
H_0 & = (C - 4D) \sigma_0 + (M - 4B) \sigma_z, \nonumber \\
V_{x} & = D \sigma_0 + B \sigma_z + \frac{1}{2i} A \sigma_x, \\
V_{y} & = D \sigma_0 + B \sigma_z - \frac{1}{2i} A \sigma_y. \nonumber
\end{align}

Since the interesting physics occurs at the boundary (topological insulators are gapped in the bulk but have conducting edge states),
we consider a semi-infinite 2D sheet.

\begin{figure}
\centerline{\includegraphics[width=95mm]{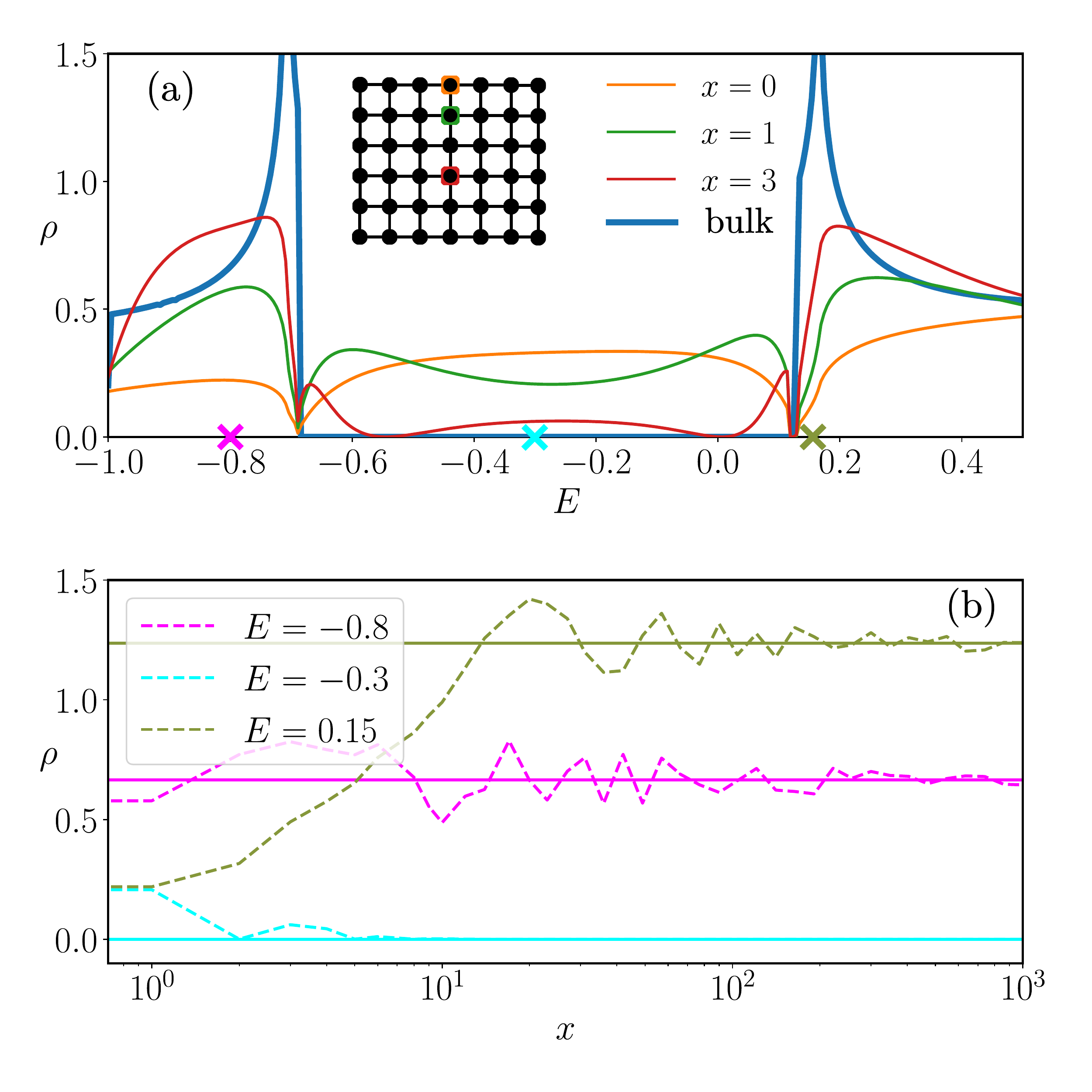}}
\caption{The quantum spin Hall model on a semi-infinite plane. (a) LDOS of the quantum spin Hall in function of energy for a bulk system (infinite in the $x$- and $y$-directions) in blue and for a semi-infinite system (semi-infinite in the $x$-direction) at a position $x=0, 1$ and 3 from the boundary, as pictured by the colored dots in the inset. The crosses on the $E$-axis correspond to the energies plotted in (b). (b) LDOS of a semi-infinite system (dashed lines) in function of the distance to the boundary. The horizontal lines represent the LDOS of a bulk system.}
\label{fig:QSH}
\end{figure}

We consider a sample made of a pristine semi-infinite sheet of this topological material, as shown in the inset of Fig.~\ref{fig:QSH}a .
Fig.~\ref{fig:QSH}a shows the local density of states (LDOS) as a function of energy $E$, $\rho(x, E) =  - \text{Tr}\ \  \text{Im}\bra{\bo{r}} G(E) \ket{\bo{r}}/\pi$ at three different distances from the edge.
We also plot the bulk density of states (DOS) of the infinite sheet (i.e.\ the value of the LDOS infinitely far from the edge).
The latter shows a vanishing DOS inside the gap $[-0.7,0.7]$ of the system. The non-zero values of the LDOS inside the gap for the semi-infinite sheet is due to the presence of the propagating edge states.
Note that all quantities here are independant of $y$ as all systems are invariant by translation along that direction.
As the distance from the boundary is increased, the edge states amplitude decrease exponentially which leads to a corresponding decrease of the LDOS inside the gap.

Fig.~\ref{fig:QSH}b shows a calculation of the LDOS versus $x$ for the three energies marked by crosses in Fig.~\ref{fig:QSH}a.
We observe a slow (algebraic) convergence to the (independently calculated) bulk limit
for the two values of energy inside the spectrum, characteristic of the Friedel oscillations created by the presence of the edge.
Inside the gap, we observe a much faster exponential convergence towards zero due to the exponential localization of the edge states.
We emphasize again that the computational time to obtain this data is totally independent of the value of $x$ used in the calculation, in sharp contrast with standard approaches.

\subsection{Graphene nanoribbon} \label{sec:2d_leads}
\begin{figure*}
\centerline{\includegraphics[width=180mm]{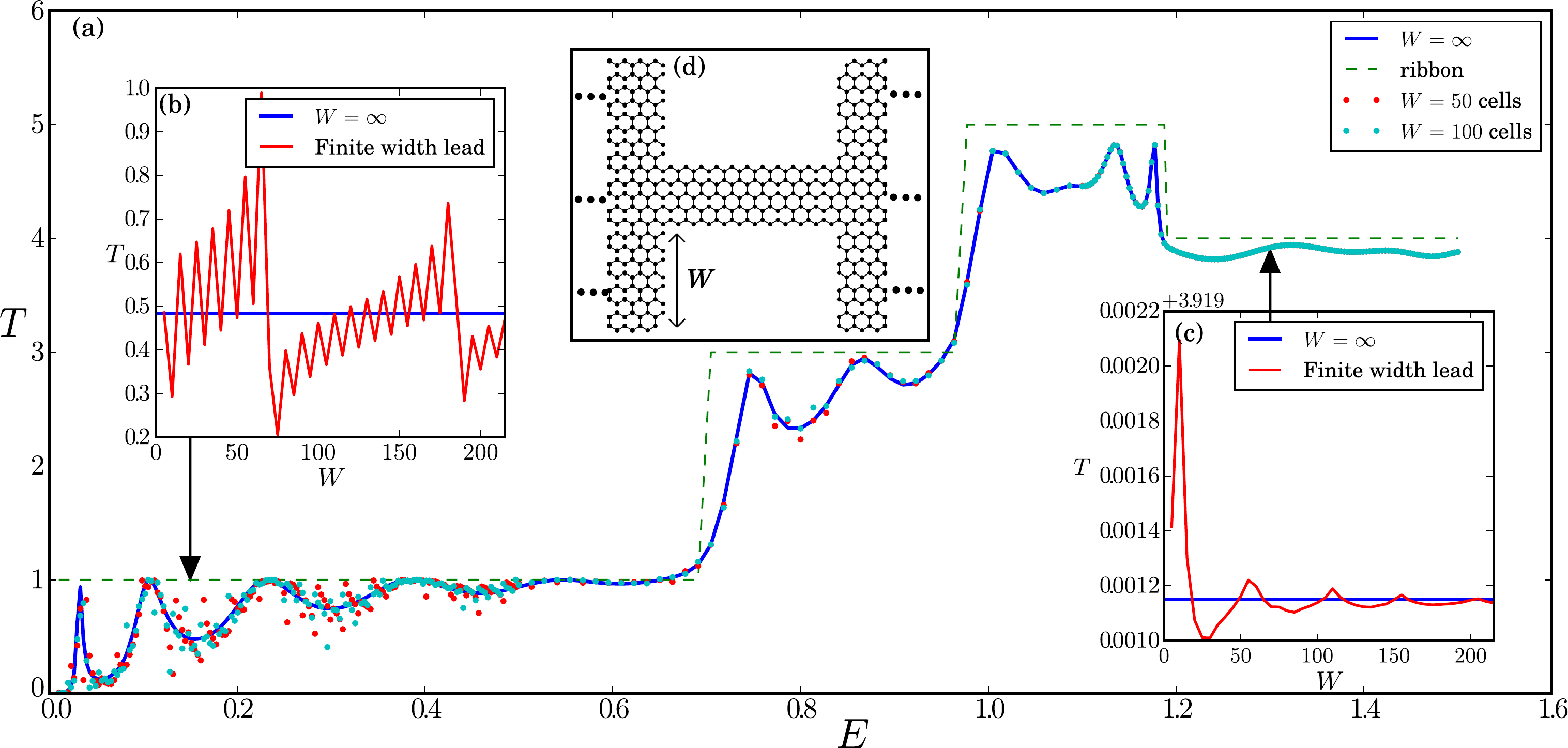}}
\caption{Graphene nanoribbon connected to two semi-infinite 2D electrodes (solid blue line) or to two finite-width quasi-1D electrodes (dots). Main panel (a): Transmission $T(E)$ versus energy $E$. The dashed line indicates the $W=0$ situation. (b) Transmission versus width $W$ for the fixed value of the energy $E=0.15$ indicated by the black arrow. The horizontal blue line shows the MTIS calculation $W=\infty$. (c) Same as (b) for $E=1.3$. (d) Schematic of the system. The electrodes extend to infinity on the right and on the left. In the $W = \infty$ case (MTIS), the leads elso extend to infinity towards the bottom and the top.}
\label{fig:2D_leads}
\end{figure*}

We now turn to a quantum transport problem:
the calculation of the conductance of a graphene nanoribbon connected to two semi-infinite graphene sheets.
The sample geometry corresponds to the one shown in Fig.~\ref{fig:2D_leads}d with the width $W$ approaching infinity.
For simplicity, we use the standard nearest-neighbor Hamiltonian on a honeycomb lattice,
\begin{equation}
\hat{H}_{\rm G} = \sum_{\langle i, j \rangle} \ket{i} \bra{j},
\end{equation}
but the method applies to any other tight-binding model as well.

Obtaining the conductance of this setup with the help of a traditional quantum transport code (that is limited to quasi-1D leads of finite width) involves a series of calculculations where the semi-infinite sheets are approximated by leads of increasing finite width.
The conductance in the thermodynamic limit $W=\infty$ is then extrapolated from these finite-size results.
In contrast, our new method works directly in the thermodynamic limit.

We note that this particular system could also be addressed with the technique of Ref.~\onlinecite{Fabry_perot_graphene}
that uses an iterative Lanczos-like approach that takes advantage of the existence of the nanoribbon constriction to restrict the Hamiltonian to the states that are connected to it.
This approach is orthogonal (and possibly complementary) to our method which exploits translational invariance.

In order to directly compute observables of the infinitely wide sample, we first calculate the Green's functions of the two semi-infinite 2D graphene sheets and then proceed to connect them to the central nanoribbon using the Dyson equation.
Following the standard formalism~\cite{datta_1995}, the differential conductance $g_{ll'}= dI_l/dV_{l'}$ between electrode $l'$ and $l$ is then given in terms of the total transmission probability $g_{ll'} = (e^2/h) T_{ll'}$ with
\begin{equation}
T_{ll'} = \operatorname{Tr}(\hat G \Gamma_l \hat G^\dagger \Gamma_{l'}),
\label{eq:Transmission}
\end{equation}
where $\hat G$ designates the Green's function of the total system,
and
\begin{equation}
\Gamma_l = i(\Sigma_l - \Sigma_l^\dagger)
\end{equation}
the rate matrix expressed in terms of the lead self-energy
\begin{equation}
\Sigma_l = W_{l} G_{l} W^\dagger_{l}.
\end{equation}
In the above definition, $G_l$ is the surface Green's function of one graphene electrode (here a semi-infinite 2D graphene sheet).
$W_{l}$ corresponds to the matrix elements of the Hamiltonian that connect the central ribbon to electrode $l$.

Fig.~\ref{fig:2D_leads} shows the results of a calculation for a particular geometry of the central graphene ribbon (inset d).
The main panel displays the transmission probability $T_{ll'}$ between the two leads as a function of energy.
We observe that the results of finite-width computations (colored dots) are scattered around the infinite-width values (solid line) that have been obtained with the MTIS method.

For perspective, the dashed line represents the result of a calculation for $W=0$,
i.e. for a simple infinite ribbon of fixed width.
The conductance is quantized and the total transmission simply counts the number of conducting sub-bands at the Fermi level.
In contrast,  we note the presence of fluctuations in the other curves.
They are caused by the abrupt widening on both ends of the ribbon that induces reflexions which in turn create a Fabry-Perot cavity.\cite{Fabry_perot_graphene} The associated interference pattern is at the origin of the fluctuations.

At lower energies, we observe a clear difference between the finite-width data and the infinite-width case.
This is to be expected since, for small energies, the associated wave length $\lambda \sim 1/E$ is comparable to the ribbon width
and hence finite size effects occur.
The apparent noise in the finite-width data is due to rapid oscillations in energy.
Fig.~\ref{fig:2D_leads}b shows the convergence
of the result as a function of the width for a fixed energy.
We find the convergence towards infinite-width limit (blue horizontal curve) slow and oscillatory.
Consequently, accessing the large-width physics with standard finite-width techniques turns out to be extremely difficult at low energies.

For large values of the energy (typically $E > 1$), we find that the finite-width data
is almost indistinguishable from the $W=\infty$ curve.
A plot of the convergence at
$E=1.3$ is shown in Fig.~\ref{fig:2D_leads}c.
In this regime, the MTIS approach has nevertheless a clear computational advantage because the computational cost of the finite-width quasi-1D electrodes scales as $W^3$.
With our current implementation, we find that computing directly the $W=\infty$ limit is faster than the finite-width calculation as soon as $W \ge 100$ for $E = 0.5$.

\subsection{Weyl semimetal three-terminal device} \label{sec:Weyl}
\begin{figure*}
\centerline{\includegraphics[width=180mm]{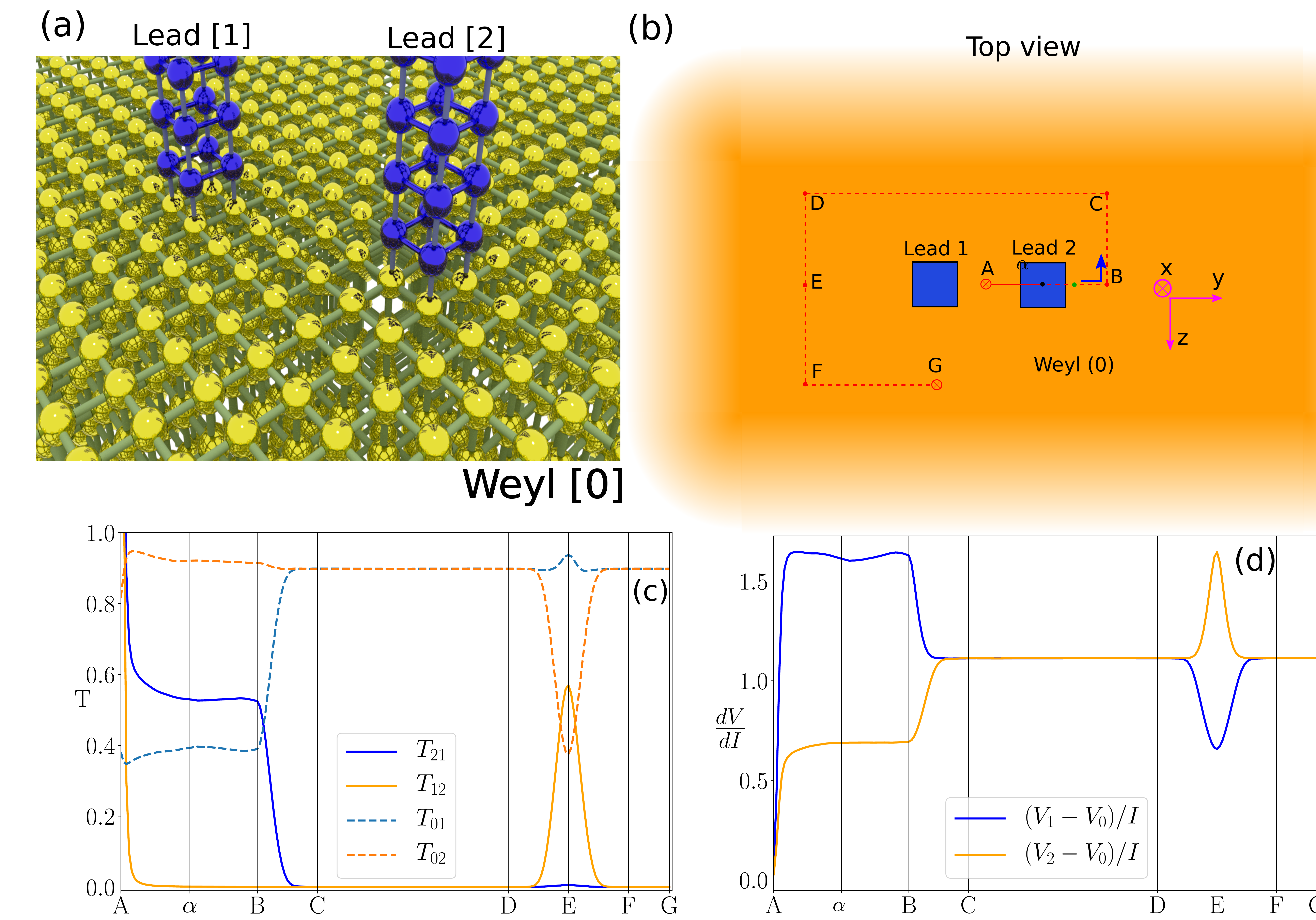}}
\caption{Weyl semimetal three-terminal device. (a) 3D view. Two squared one-dimensional leads ($5\times5$ sites in the simulations) in blue are connected to the surface of a 3D semi-infinite Weyl semimetal (in yellow). The Weyl semimetal part spans the $yz$-plane for all $x \leq 0$. (b) Same as (a) but cut perpendicularly from the $z$-axis. The dashed line indicates the trajectory followed by lead 2. The green point labeled by $\alpha$ indicates the position of lead 2 in the plot of Fig.~\ref{fig:Weyl_energy}. (c) Transmissions in between the leads (0 is the Weyl semimetal, 1 and 2 the two blue leads) at $E=0.02$ as the right lead is moved from $A$, where the two blue leads are in contact, to G following the path shown in (a). The distance AB correspond to 55 sites and BC to 25 sites. (d) Differential resistance plotted along the path shown in (a), where $V_1$ and $V_2$ are the tensions in the two leads. The Weyl part is connected to the ground, such that the current can only come in or out in the 1D leads.}
\label{fig:Weyl_moving_leads}
\end{figure*}

We conclude this section with an application that showcases the full power of the MTIS approach.
We consider a 3D topological Weyl semi-metal with impurities on its surface and calculate the differential conductance of a three-terminal geometry.
This computaton combines the difficulties of addressing an infinite 3D system (half of the space is filled with the 3D material) with disorder, in a multiterminal geometry and in presence of surface states (the so-called Fermi arcs).
The geometry of the device is shown in Fig.~\ref{fig:Weyl_moving_leads}a with a top view shown in \ref{fig:Weyl_moving_leads}b.

We use a 3D 4-band model of a Weyl semimetal that is defined as \cite{Vazifeh2013Electromagnetic}

\begin{equation}
\begin{split}
\mathcal{H}_{\rm WEYL}(\bo{k}) = \tau_z \left[ t (\sigma_x \sin(k_x) + \sigma_y \sin(k_y)) + t_z \sigma_z \sin(k_z)\right]\\
+ \mu(\bo{k}) \tau_x \sigma_0 + \frac{1}{2} b_0 \tau_z \sigma_0 + \frac{1}{2} \beta \tau_0 \sigma_z,
\end{split}
\end{equation}
with
\begin{equation*}
\mu(\textbf{k}) = \mu_0 + t(2 - \cos(k_x) - \cos(k_y)) \\ + t_z (1 - \cos(k_z))).
\end{equation*}
The Pauli matrices $\tau_i$ and $\sigma_i$ ($i=x, y, z$) act, respectively, on the orbital and spin degrees of freedom.
We use the parameters $t = 2$, $t_z = 1$, $\mu_0 = -0.1$, $b_0 = 0$ and $\beta = 1$. After discretization, the on-site matrix $H_0$ and hopping matrices $V_x$, $V_y$ and $V_z$ are given by
\begin{align}
H_0 & = (\mu_0 + 2t + t_z) \tau_x + \frac{1}{2} b_0 \tau_z + \frac{1}{2} \beta \sigma_z,\\ \nonumber
V_{x} & = \frac{1}{2}i t \tau_z \sigma_x - \frac{1}{2} t \tau_x \sigma_0, \\ \nonumber
V_{y} & = \frac{1}{2}i t \tau_z \sigma_y - \frac{1}{2} t \tau_x \sigma_0, \\ \nonumber
V_{z} & = \frac{1}{2}i t_z \tau_z \sigma_z - \frac{1}{2} t_z \tau_x \sigma_0.
\end{align}

\begin{figure}
\centerline{\includegraphics[width=88mm]{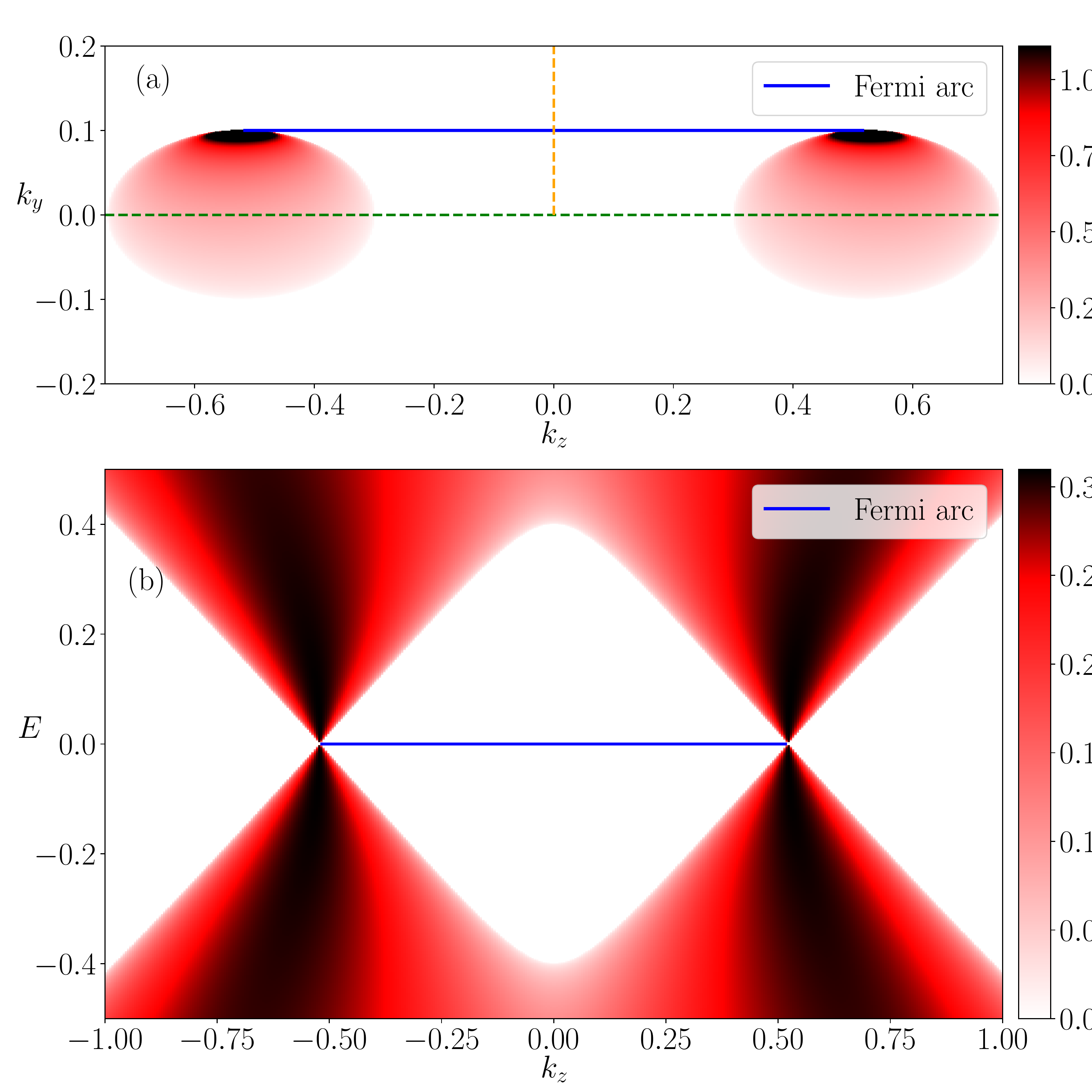}}
\caption{LDOS of a semi-infinite Weyl semimetal. (a) LDOS of the semi-infinite system as a function of $k_y$ and $k_z$, at $E=0.2$. The blue line connecting the cones is a bound state, the so-called Fermi arc. (b) LDOS of the semi-infinite system as a function of $k_z$ and $E$, at $k_y = 0$.}
\label{fig:LDOS_Weyl}
\end{figure}

\begin{figure}
\centerline{\includegraphics[width=93mm]{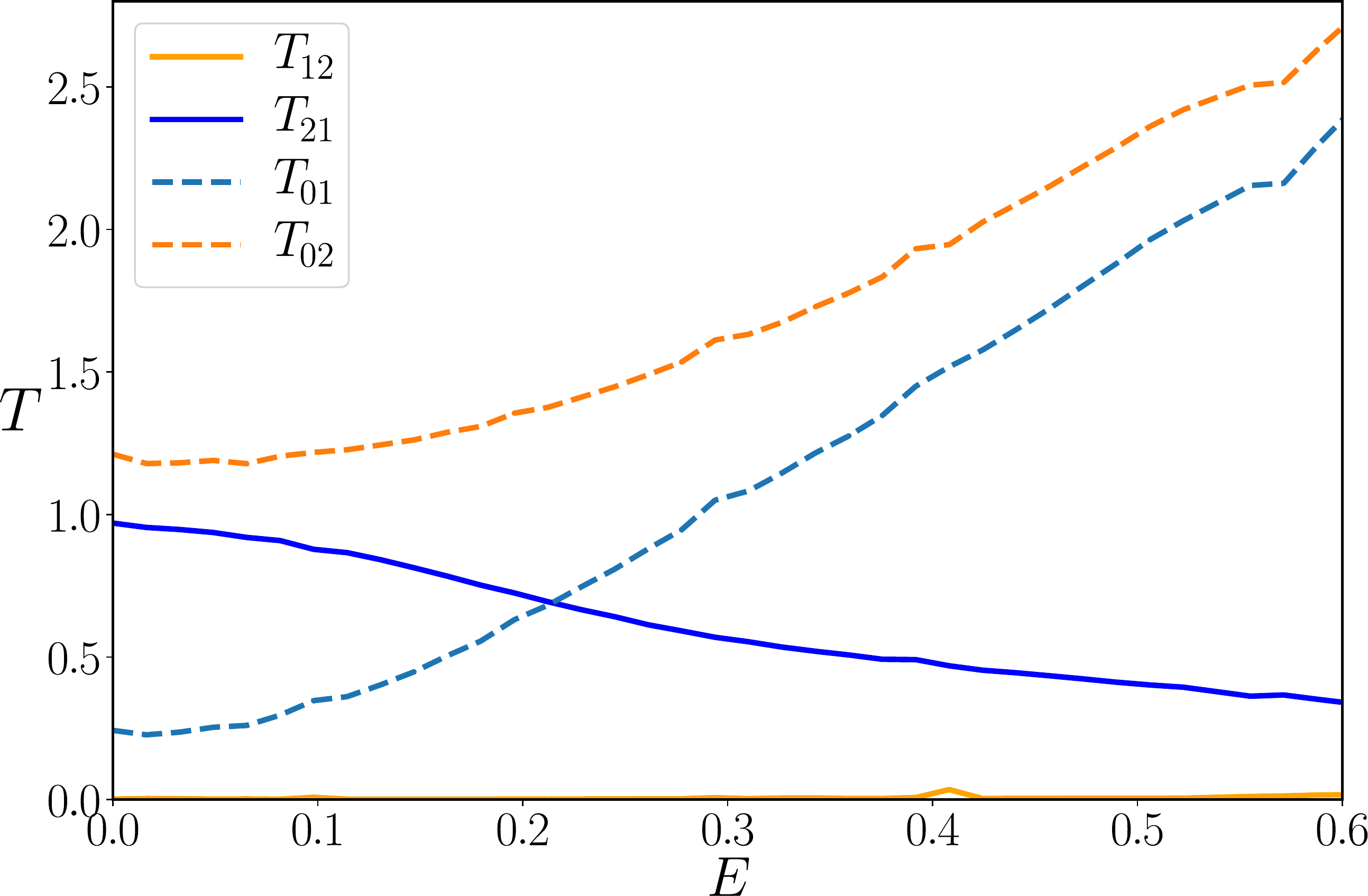}}
\caption{Transmission between the two leads as function of the energy. The leads contain 25 sites and are similar to simulations shown in Fig.~\ref{fig:Weyl_moving_leads}c and d. Lead 2 is located 25 sites away from lead 1, as indicated by the point $\alpha$ in Fig.~\ref{fig:Weyl_moving_leads}b, c and d. This length is sufficient to reach the plateau of $T_{21}$ between points A and B.}
\label{fig:Weyl_energy}
\end{figure}

\begin{figure*}
\centerline{\includegraphics[width=180mm]{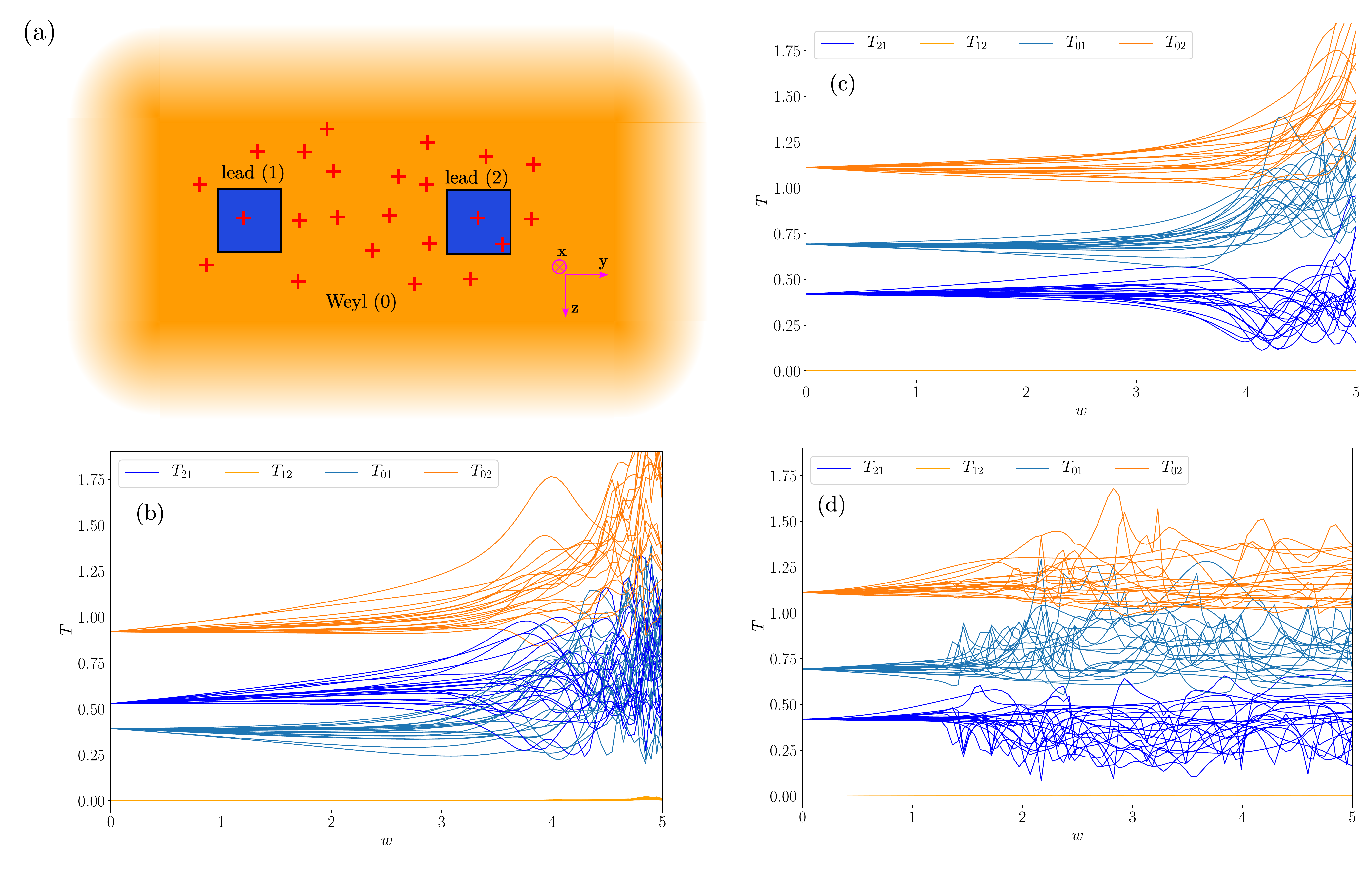}}
\caption{ Transmission of the disordered Weyl surface. (a) The two leads are placed on the surface at $r_1 = (0, -20, 0)$ and $r_2 = (0, 20, 0)$. 30 impurities are scattered randomly on the surface around the leads in the rectangle $-6 < z < 6$ and $-30 < y < 30$, so that the impurity density is $5\%$ in that rectangle. The curves in Fig.~\ref{fig:1D_integrand} and Fig.~\ref{fig:1D_principal_value} are matrix elements of the GF respectively plotted along the green or orange dotted lines. (b) Transmissions between the leads and the bulk as a function of the strength of the impurity at $E=0.02$. Same color correspond to the same transmissions for 20 different samples (distribution of impurities). (c) Same as (b), for $E=0.2$, (d) Same as (c), but a random impurity matrix is chosen on each site.}
\label{fig:Weyl_impurities}
\end{figure*}

The geometry of Fig.~\ref{fig:Weyl_moving_leads}a consists of the Weyl semimetal occupying the $x \geqslant 0$ half of the 3D space to which two quasi-1D electrodes are attached.
We also deposit some impurities on the surface. The complex sequence of submethods used to calculate the differential conductance of this geometry is shown in Fig.~\ref{fig:system_flowchart}.
The quasi-1D electrodes are considered to be made from a normal metal described, respectively, by on-site and hopping parameters $t=1$ and $\mu = 0$.
The electrodes have a square cross section of $5\times 5$ sites.
In the following, the quasi-1D electrodes are labeled as 1 and 2, and the infinite Weyl semi-metal is labeled as 0.
The conductance matrix that relates the current $I_a$ to the voltage $V_b$ applied to lead $b$ takes the form
\begin{equation}
\begin{pmatrix}
I_0 \\
I_1 \\
I_2
\end{pmatrix}
 =
 \frac{e^2}{h}
 \begin{pmatrix}
T_{02} + T_{01} & -T_{01} & -T_{02} \\
 -T_{10} & N_{1} - R_{11} & -T_{12} \\
 -T_{20} & -T_{21} & N_{2} - R_{22}
 \end{pmatrix}
\begin{pmatrix}
V_0 \\
V_1 \\
V_2
\end{pmatrix},
\label{eq:conductance_matrix}
\end{equation}
where $N_{a} = \sum_j T_{lj} = \sum_j T_{jl}$ is the number of channels in the quasi-1D electrode $a$. The conductance matrix conserves current and is ``gauge invariant'' in the B\"uttiker sense, i.e.\ raising simultaneously all voltages by the same amount does not change the current\cite{Buttiker1995}.
Notice the special treatment of electrode $0$: since $R_{00}$ and $N_{0}$ are both infinite,
only their difference is a well-defined quantity.
The total reflection terms $R_{aa}$ are given by\cite{Wimmer_thesis}
\begin{equation}
\begin{split}
  R_{aa} = { } & N_{\rm a} + \text{Tr}(\Gamma_a \hat G \Gamma_a \hat G^\dagger) \\
  & + i \big( \text{Tr}(\Gamma_a \hat G) - \text{Tr}(\Gamma_a \hat G^\dagger) \big),
\end{split}
\end{equation}
where $G_a$ is the Green's function of the full system at the surface of a quasi-1D electrode.

We now present the results of the calculation for a clean surface.
We assume the geometry
of Fig.~\ref{fig:Weyl_moving_leads}b where one electrode (serving as Ohmic contact) is fixed on the surface while the second one scans the surface modeling the tip of a STM (scanning tunneling microscope).
The letters ABCDEFG indicate the trajectory of the STM tip on the surface.
Fig.~\ref{fig:Weyl_moving_leads}c shows the transmissions between the different electrodes as a function of the position of the STM tip on its trajectory for a value of energy close to the Weyl points.
At this energy, the bulk density of states is very small and transport is entirely dominated by the surface states, the so-called Fermi arcs. These arcs are extremely
anisotropic with a dispersion relation $E = v k_y$ so that they only propagate along the $y$ axis in the {\it positive} direction.
It follows that transport between electrodes $1$ and $2$ is only possible when they are aligned along the $y$-direction; and only from $1$ to $2$, not from $2$ to $1$. This is illustrated by the curves between A and B where $T_{21}\approx 0.55$ and $T_{12} \approx 0$.
As soon as this alignment is lost, both $T_{21}$ and $T_{12}$
vanish and all the current disappears into the substrate.
Note that the above results are closely linked to the band structure of the model considered here which only contains one pair of Weyl points.
In real materials, there might be several pairs of such points\cite{xu2015discovery} and surface reconstruction can also lead to a bending of the Fermi arcs\cite{deng2016experimental} which would lead to a decrease of the extreme anisotropy observed in our numerics.

Intermediate steps in the calculation of the above transport curve involve the computation of the Fermi arcs wavefunction, first in momentum space and then in real space.
This information is necessary to analyze the final calculation of observables.
The color plot of Fig.~\ref{fig:LDOS_Weyl} shows the local density of states of the pristine surface,
\begin{equation}
\rho(x,k_y,k_z E) = - \frac{1}{\pi} \text{Tr} \ \text{Im}\langle x,k_y,k_z|G(E)|x,k_y,k_z\rangle,
\end{equation}
while the blue line indicates the position of the Fermi arc.
From such plots, one can extract for example the dispersion relation of the Fermi arcs.

Fig.~\ref{fig:Weyl_energy} shows the different transmissions as a function of energy $E$ when the STM tip is positioned at point A'. The transmission towards the bulk increases quadratically which is consistent with a linear density of states in the bulk (dashed lines).
More interestingly, the transmissions $T_{21}$ and $T_{12}$ are very weakly affected by the presence of the bulk states and remain very anisotropic ($T_{12}\approx 0$). Hence, transport on the surface is very weakly affected by the presence of the bulk states and remains dominated by the Fermi arcs\cite{PhysRevB.93.235127}.

Fig.~\ref{fig:Weyl_moving_leads}d shows the differential conductance measured across the two
top contacts in the following setup: one injects a current $I$ into electrode $1$ and retrieves the same current from electrode $2$ (i.e. vanishing current leaves electrode 0).
The strong anisotropy observed in the two non-local resistances $(V_1-V_0)/I$ and $(V_2-V_0)/I$ bears the signature of the Fermi arc anisotropy.

We now turn to the study of the resilience of the above picture in the presence of disorder.
While disorder in bulk Weyl semimetals has been the focus of several studies\cite{PhysRevLett.113.026602, annurev-conmatphys-033117-054037, PhysRevLett.115.076601, PhysRevLett.115.246603},
the effect on the surface has received much less attention\cite{Das_sarma_disorder_weyl}. We consider discrete impurities randomly scattered at the surface in the region around the leads (see Fig.~\ref{fig:Weyl_impurities}a).

We model the impurities by modifying $5\%$ of the on-site Hamiltonian matrices located around the leads (20 to 30 sites) to which we add a shift in energy $w \cdot \mathbb{1}_4$ (Fig.~\ref{fig:Weyl_impurities}b and c
or a fully random on-site matrix $w \sum_{i, j} h_{i, j} \sigma_i \tau_j $ where $h_{i, j}$ are random numbers from the $[-1, 1]$ interval (Fig.~\ref{fig:Weyl_impurities}d).
Different traces correspond to different samples.

In the low energy regime of Fig.~\ref{fig:Weyl_impurities}b, at energy $0.02$, the effect of the impurities is only significant for a large value of $w$ comparable to the bandwidth of the model (equal to approximately $5$), which indicates that the anisotropic transport on the surface is resistant to the presence of disorder. This is not unexpected since at this energy the density of bulk states to scatter to is very low.
However, the same situation is shown in Fig.~\ref{fig:Weyl_impurities}b for an energy ten times larger, $E=0.2$.
The resilience to disorder persists when we use the random matrix disorder of Fig.~\ref{fig:Weyl_impurities}d that breaks all possible symmetries.
We conclude that the topological protection of the Fermi arcs can be observed directly in the differential conductance which is not perturbed by the presence of the bulk states.
Our observations are compatible with the statement that disorder only renormalizes the dispersion relation\cite{Das_sarma_disorder_weyl},
which remains non-zero even close to Anderson localization.

\section{The Residue problem} \label{sec:1D_problem}
We turn to the detailed derivation of the formalism and algorithms.
In a first step we calculate a finite set of matrix elements for the Green's function of a TIS, i.e. $\bra{\bo{r'},\mu }\hat g_l(E) \ket{\bo{r},\nu}$
for a set of values $(\bo{r'},\mu,\bo{r},\nu)$. The Green's function of the TIS is formally defined as the inverse of the
Hamiltonian (shifted in energy):
\begin{equation}
\hat g_l(E) = \frac{1}{E - \hat H_l^{\infty} + i \eta}.
\label{eq:bulk_gf_def2}
\end{equation}
Since the TIS is invariant by translation, $\hat H_l^{\infty}$ can be diagonalized in momentum space.
The momentum states $|\alpha \bo{k}\rangle$ satisfy $\hat H_l^{\infty}|\alpha \bo{k}\rangle=E|\alpha \bo{k}\rangle$ and take the form
\begin{equation}
|\alpha \bo{k}\rangle = \sum_{\bo{r},\mu} \Psi_{\alpha\bo{k}}(\mu) e^{i\bo{k}.\bo{r}} |\bo{r},\mu\rangle,
\end{equation}
where the eigenvectors $\Psi_{\alpha\bo{k}}$ are eigenstates
\begin{equation}
\mathcal{H}(k)\Psi_{\alpha\bo{k}}=E(\bo{k})\Psi_{\alpha\bo{k}}
\end{equation}
of the momentum Hamiltonian
\begin{equation}
\begin{split}
\mathcal{H}(k) \equiv {} & H_0 + V_x^{\dagger} e^{ik_x} + V_x e^{-ik_x} \\
& + V_y^{\dagger} e^{ik_y} + V_y e^{-ik_y}+ V_z^{\dagger} e^{ik_z} + V_z e^{-ik_z}.
\end{split}
\end{equation}
We arrive at
\begin{multline}
  \bra{\bo{r'},\mu }\hat g_l(E) \ket{\bo{r},\nu} = \\
\sum_\alpha\int_{-\pi}^{+\pi} \frac{d^3 \bo{k}}{(2\pi)^3} e^{i\bo{k.(r - r')}}
\frac{\Psi_{\alpha\bo{k}}(\mu)\Psi_{\alpha\bo{k}}^*(\nu)}{E-E(\bo{k}) +i\eta}.
\label{eq:bulk_gf_ft2}
\end{multline}
Eq.~\eqref{eq:bulk_gf_ft2} forms the starting point of our calculation. The momentum Hamiltonian $\mathcal{H}(k)$
is a small $N_o\times N_o$ matrix that can easily be diagonalized numerically. Hence, at first glance, the Fourrier transform
in Eq.~\eqref{eq:bulk_gf_ft2} could be performed numerically using stanrd quadrature rules such as Simpson. Such an approach is too naive however,
since the fraction $1/[E-E(\bo{k}) +i\eta]$
diverges when $E(\bo{k})$ crosses $E$. As $\lim_{\eta\rightarrow 0} 1/(X+i\eta)$ equals $P(1/X) -i\pi \delta(X)$, i.e.\ the principal part and a Dirac distribution,
the integral over at least one of the momentum variables (here $k_x$) must be calculated analytically.
This is the residue problem which we discuss in this section.

For a fixed value of $k_y$ and $k_z$, we are left with a 1D TIS problem described by an on-site matrix $H$ and a hopping matrix $V$:
\begin{align}
H & = H_0 + V_y^{\dagger} e^{ik_y} + V_y e^{-ik_y}+ V_z^{\dagger} e^{ik_z} + V_z e^{-ik_z} \\
V & = V_x. \nonumber
\end{align}

\subsection{Formulation of the residue problem} \label{sec:Bloch_theorem}

The 1D translationally invariant Hamiltonian $\hat{H}_{\rm \mrm{1D}}$ of the system,
\begin{equation}
 \hat{H}_{\rm \mrm{1D}} =
 \begin{pmatrix}
\ddots & \ddots & \ddots &  \\
& V & H & V^{\dagger} &  \\
& & V & H & V^{\dagger} & \\
& & & V & H & V^{\dagger} &\\
& & & & \ddots & \ddots & \ddots \\
 \end{pmatrix},
\label{eq:total_hamiltonian}
\end{equation}
consists of repeated blocks as in Fig.~\ref{fig:system}a,b. Each block is described by $N_{o} \times N_{o}$ matrices $H$ (on-site) and $V$ (hopping to the next cell). We seek a finite number of elements of the 1D Green's function,
\begin{equation}
 \hat{g}^\mrm{1D}(E) = \lim_{\eta \to 0^+}\frac{1}{E - \hat{H}_{\rm \mrm{1D}} + i \eta}.
 \label{eq:GF_def}
\end{equation}
Introducing the $N_{o} \times N_{o}$ matrix $g^\mrm{1D}_{x, x'}(E)$ whose elements are given by
\begin{equation}
[g^\mrm{1D}_{x, x'}(E)]_{\mu\nu} \equiv \bra{x,\mu }\hat{g}^\mrm{1D}\ket{x',\nu },
\end{equation}
the residue problem consists of calculating the integral
\begin{equation}
\label{eq:g_1D_eigenvector_expansion}
 g^\mrm{1D}_{x, x'}(E) = \sum_\alpha \int_{-\pi}^{\pi} \frac{dk}{2 \pi} \frac{e^{ik(x - x')} \Psi_{\alpha k} \Psi_{\alpha k}^\dagger}{E - E(k) + i \eta},
\end{equation}
or, in a more compact form,
\begin{equation}
\label{eq:g_1D_real_axis}
g^\mrm{1D}_{x, x'}(E) = \int_{-\pi}^{\pi} \frac{dk}{2 \pi} \frac{e^{ik(x - x')} }{E - \mathcal{H}(k) + i \eta},
\end{equation}
with
\begin{equation}
\mathcal{H}(k) \equiv H + V^{\dagger} e^{ik} + V e^{-ik}.
\end{equation}

\subsection{Solution to the residue problem} \label{sec:Residue_theorem}
Here we provide without justification the solution of the residue problem.
The proof of this result is given in the following section.

The  first step of the algorithm is to solve the generalized eigenvalue problem
\begin{equation}
[E - \mathcal{H}(k) ] \phi = 0,
\label{eq:gen_pb}
\end{equation}
where $E$ is an {\it input} and we seek the values of $k$ for which the above equation has a solution $\phi$. We emphasize
that this is a very different problem from diagonalizing $\mathcal{H}(k)$ for a given value of $k$.
Introducing explicitly the form of $\mathcal{H}(k)$ and designating $e^{ik}$ as $\lambda$,
Eq.~\eqref{eq:gen_pb} takes the form
\begin{equation}
\left(
\begin{matrix}
H-E & V^\dagger \\
1 & 0 \\
\end{matrix}\right)
\left(
\begin{matrix}
\phi \\
\xi \\
\end{matrix}\right)
=\frac{1}{\lambda}
\left(
\begin{matrix}
-V & 0 \\
0 & 1 \\
\end{matrix}\right)
\left(
\begin{matrix}
\phi \\
\xi \\
\end{matrix}\right),
\end{equation}
which is a generalized eigenproblem.
Solving this problem numerically can be handled by the
Kwant numerical package\cite{kwant_article} for example.

The second step is to sort the resulting set of eigenvectors/eigenvalues $(\phi_a,\lambda_a)$ into two categories:
{\it left}-goers and {\it right}-goers. Eigenstates with $\lambda_a <1$ ($\lambda_a >1$) are attributed to the right- (left-) goers category.
Eigenstates with
$|\lambda_a|=1$ are classified according to the velocity
\begin{equation}
v_a = \phi^\dagger_a V^\dagger e^{ik_a} - V e^{-ik_a} \phi_a.
\end{equation}
States with positive (negative) velocity are right- (left-) goers.

The third step is to construct the projectors onto the different subspaces
spanned by the eigenvectors $\phi_a$:
\begin{equation}
P_\lambda = \sum_{\lambda_a = \lambda} \phi_a \phi_a^\dagger.
\end{equation}
The special case of projector $P_0$ corresponds to the projector onto the kernel of $V$
($VP_0 = 0$). Introducing the matrix
\begin{equation}
\partial_k\mathcal{H}(k) = i (V^{\dagger} e^{ik} - V e^{-ik}),
\end{equation}
the result of the calculation reads
\begin{align}
 g^\mrm{1D}_x(E) = & \sum_{\lambda} i \lambda^x P_\lambda \frac{1}{E - \mathcal{H}(\lambda) - \partial_k\mathcal{H}(\lambda) P_\lambda} \nonumber \\
 & + \delta_{x,0} P_0 \frac{1}{[(E - H)P_0 + V]},
 \label{eq:chief}
\end{align}
where the sum is extended to the right-goers for $x\ge 0$ and to the left-goers for $x<0$.
Eq.~\eqref{eq:chief} is the chief analytical result
of this article. It provides a fast and stable solution to the residue problem.
The computation of a set of different values of $x$ involves only little numerics with small ($N_o\times N_o$) matrices.
The matrices in front of the $\lambda^x$ terms can be cached so that the computational effort to obtain $N_x$ different values of $x$ (for a fixed energy) scales as $N_o^4 + N_o^2 N_x$.

\subsection{Proof of Eq.~\eqref{eq:chief}}
\begin{figure}
\centerline{\includegraphics[width=80mm]{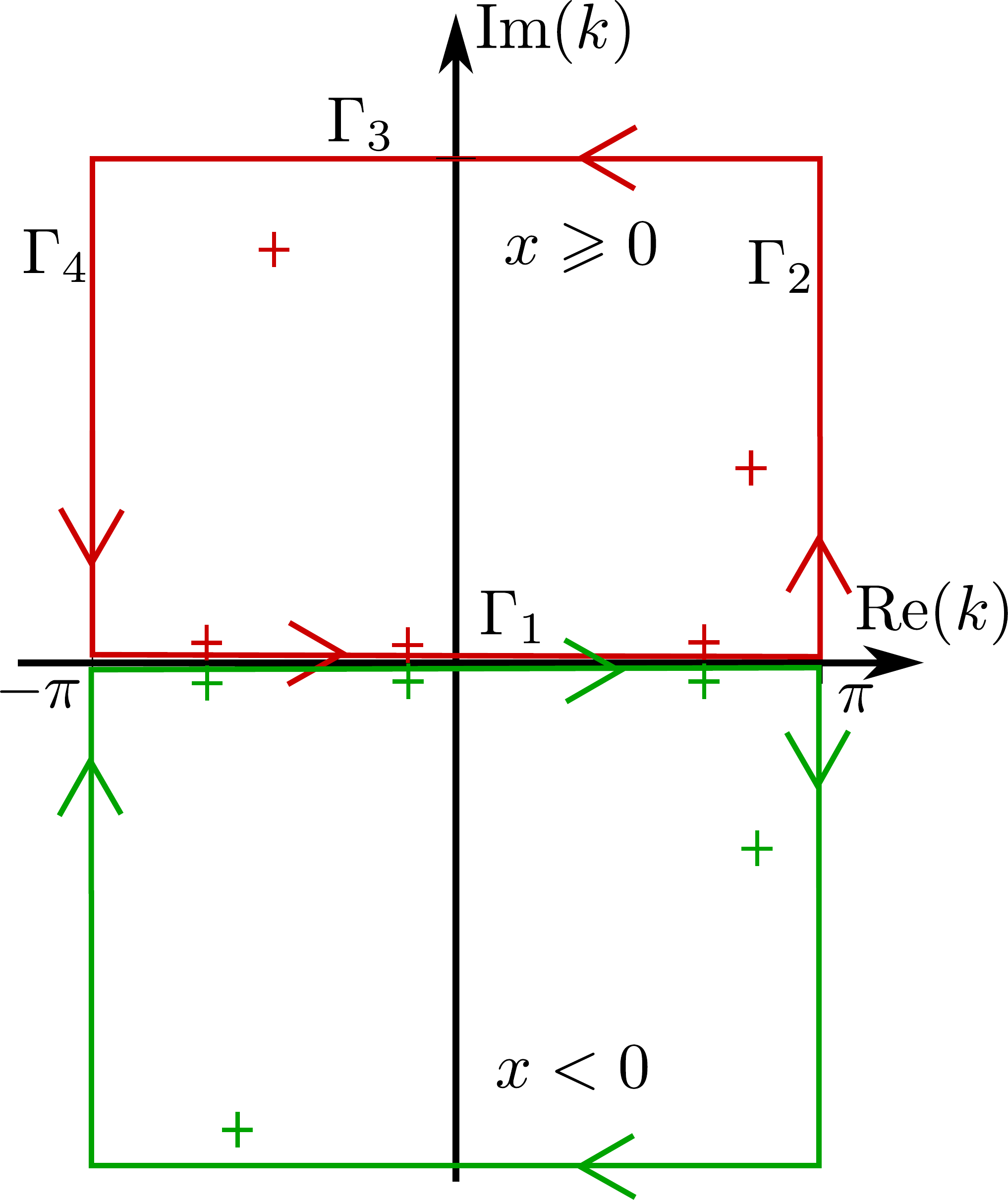}}
\caption{The two possible path integrals in the complex plane. The segments on the real axis goes from $-\pi$ to $\pi$ so it correspond to the bounds of Eq.~\eqref{eq:g_1D_real_axis}.
The crosses show the localization of the poles corresponding to solutions of Eq.~\eqref{eq:gen_pb}. The red ones correspond to right goers and the green ones to left goers. 
The integration is performed on the red (green) contour if $x \geqslant 0$ ($x < 0$). 
Each pole come in pair because for a solution $\lambda$ there is a corresponding solution $\frac{1}{\lambda*}$. The three propagative modes ($|\lambda| = 1$) are located at an infinitely small distance $\eta$ from the real axis.}
\label{fig:contour}
\end{figure}

In a first step,
we extend the integral of Eq.~\eqref{eq:g_1D_real_axis} onto the complex $k$-plane.
For $x \geqslant 0$, we use the red contour
shown in Fig.~\ref{fig:contour} which consists of four branches: the integral over $\Gamma_1$ is the orginal integral, the integrals over $\Gamma_2$
and $\Gamma_4$ compensate each other by symmetry of the integrand and the integral over $\Gamma_3$ is evaluated in the limit where the horizontal segment $\Gamma_3$
is shifted towards $i\infty$.
For $x<0$, we use the mirror-symmetric green contour.
For $x\ne 0$, the factor $\lambda^x$ in the integrand is exponentially small which makes the integral over $\Gamma_3$ to vanish in the limit $\kappa\rightarrow \infty$, where $\kappa$ is the imaginary part of the momentum.
Applying Residue theorem to this contour integral, we arrive at
\begin{multline}
g^\mrm{1D}_x(E) = \sum_\lambda {\rm Res} \left(\frac{i\lambda^x }{E - \mathcal{H}(\lambda)}\right) \\
-\lim_{\kappa\rightarrow\infty}\int_{-\pi}^{\pi} \frac{dk}{2\pi} \frac{\delta_{x,0} }{E - \mathcal{H}(k+i\kappa)},
\label{eq:res}
\end{multline}
where the first terms accounts for the residues of the integrand evaluated at its poles and the second to the integral over the $\Gamma_3$ segment.

Let us focus on the first part of this expression, the residues. The first step consist in finding the poles of the integrand of
Eq.~\eqref{eq:g_1D_real_axis}. These poles corresponds to the values $k_a$ solution of Eq.~\eqref{eq:gen_pb}. These poles are either in the upper half plane $\Im m \ k_a >0$ (right goers) or lower half plane
$\Im m \ k_a <0$ (left goers). The fate of the non-evanescent solutions with $\Im m\ k = 0$ is found by remembering the existence of the infinitely small positive $\eta$. To obtain the residues, we need to calculate the expansion of the integrand close to the pole. We start by expanding
${\cal H} (k)$ around $k_a$,
\begin{equation}
E - {\cal H}(k_a + q)= [E-{\cal H}(k_a)] - \partial_k{\cal H}(k_a) q - \partial^2_k{\cal H}(k_a) \frac{q^2}{2} + \dots,
\end{equation}
where we note that high derivatives of ${\cal H}(k_a)$ are simply related to the lower ones: $\partial_k^n{\cal H}(k_a)= -\partial^{n-2}_k{\cal H}(k_a)$
for $n\ge 3$. We now seek the expansion of $1/[E - {\cal H}(k_a + q)]$ around $q=0$. The key point to notice here is that since $[E-{\cal H}(k_a)]$ is
not invertible, this expansion has a term proportional to $1/q$ that we need to calculate to apply the residue theorem. The algebra to obtain in a systematic way the coefficients of the developpement of $1/[E - {\cal H}(k_a + q)]$ is given in Appendix \ref{App:Laurent_exp}. We obtain
\begin{equation}
\label{eq:laurent}
\frac{1}{E - \mathcal{H}(k_a+q)}= \sum_{n=-1}^{+\infty} \frac{1}{q^n} D_n(k_a),
\end{equation}
with
\begin{equation}
D_{-1}=  P_{\lambda_a} \frac{1}{E-{\cal H}(k_a)-\partial_k {\cal H}(k_a) P_{\lambda_a}}.
\end{equation}
Applying residue theorem with the above expression for the matrix residue $D_{-1}$ provides the first term of Eq.~\eqref{eq:res}. Altough we did not encounter the case in practice, it is possible that the matrix $E-{\cal H}(k_a)-\partial_k {\cal H}(k_a) P_{\lambda_a}$ above is
not invertible. In that case  the sum in Eq.~\eqref{eq:laurent} starts at $-2$. Appendix \ref{App:Laurent_exp} provides a systematic iterative construction of the residue in that case. Appendix \ref{App:Laurent_exp} also provides a general algorithm to calculate the other coefficients $D_n$
if ever needed.

Let us now calculate the second term of Eq.~\eqref{eq:res}, the integral over the $\Gamma_3$ segent of the contour which is only present when
$x=0$. When $\kappa$ is large, one gets from the definition of $\mathcal{H}(k)$,
\begin{equation}
E - \mathcal{H}(k+i\kappa) = e^{\kappa -ik} [-V  + (E-H_0) e^{-\kappa +ik} ] + \mathcal{O}(e^{-\kappa}).
\end{equation}
We need to integrate $1/[E - \mathcal{H}(k+i\kappa)]$ over $k$. If the matrix $V$ is invertible, then the second term of the development is not
necessary: $1/[E - \mathcal{H}(k+i\kappa)]\approx -e^{-\kappa + ik} V^{-1}$ and the integrand (hence the integral) vanishes in the limit $\kappa\rightarrow +\infty$. However, if $V$ is not invertible, the expansion of $1/[-V  +(E-H_0) e^{-\kappa +ik} ]$ in power of $e^{-\kappa +ik}$
has a pole $\propto e^{\kappa -ik}$ which provides a term of order $\mathcal{O}(1)$ in the integrand. Using the result of Appendix \ref{App:Laurent_exp}, we obtain
\begin{equation}
\frac{1}{-V  + (E-H_0) e^{-\kappa +ik} } =\frac{1}{e^{-\kappa +ik}} P_0 \frac{1}{(E-H)P_0-V} + \mathcal{O}(1).
\end{equation}
It follows that the integrand of the second term of Eq.~\eqref{eq:res} is $k$-independent and can be calculated explicitely. This concludes the proof of Eq.~\eqref{eq:chief}.

\section{Numerical Fourier transform} \label{sec:periodic_2D_sys}
Once the integration over $k_x$ has been performed using the complex integration technique of the previous section,
the TIS Green's function of Eq.~\eqref{eq:bulk_gf_ft2} takes the form
\begin{equation}
g^\mrm{3D}_{x,y,z}(E) =\int_{-\pi}^\pi \frac{dk_ydk_z}{(2\pi)^2} e^{i(k_y y + k_z z)} g^{\rm 1D}_x(k_y, k_z),
\label{eq:bulk_gf_ft3}
\end{equation}
where $g^{\rm 1D}_x(k_y, k_z)$ is the result of the residue solver.
In this section, we focus on the case where
the system covers the full 3D space, $-\infty \le x \le +\infty$.
To study, e.g., a surface with $-\infty \le x \le 0$, one introduces intermediate steps before the integration over $k_y$ and $k_z$.
These steps are discussed in the sections further below.

\begin{figure}
\centerline{\includegraphics[width=93mm]{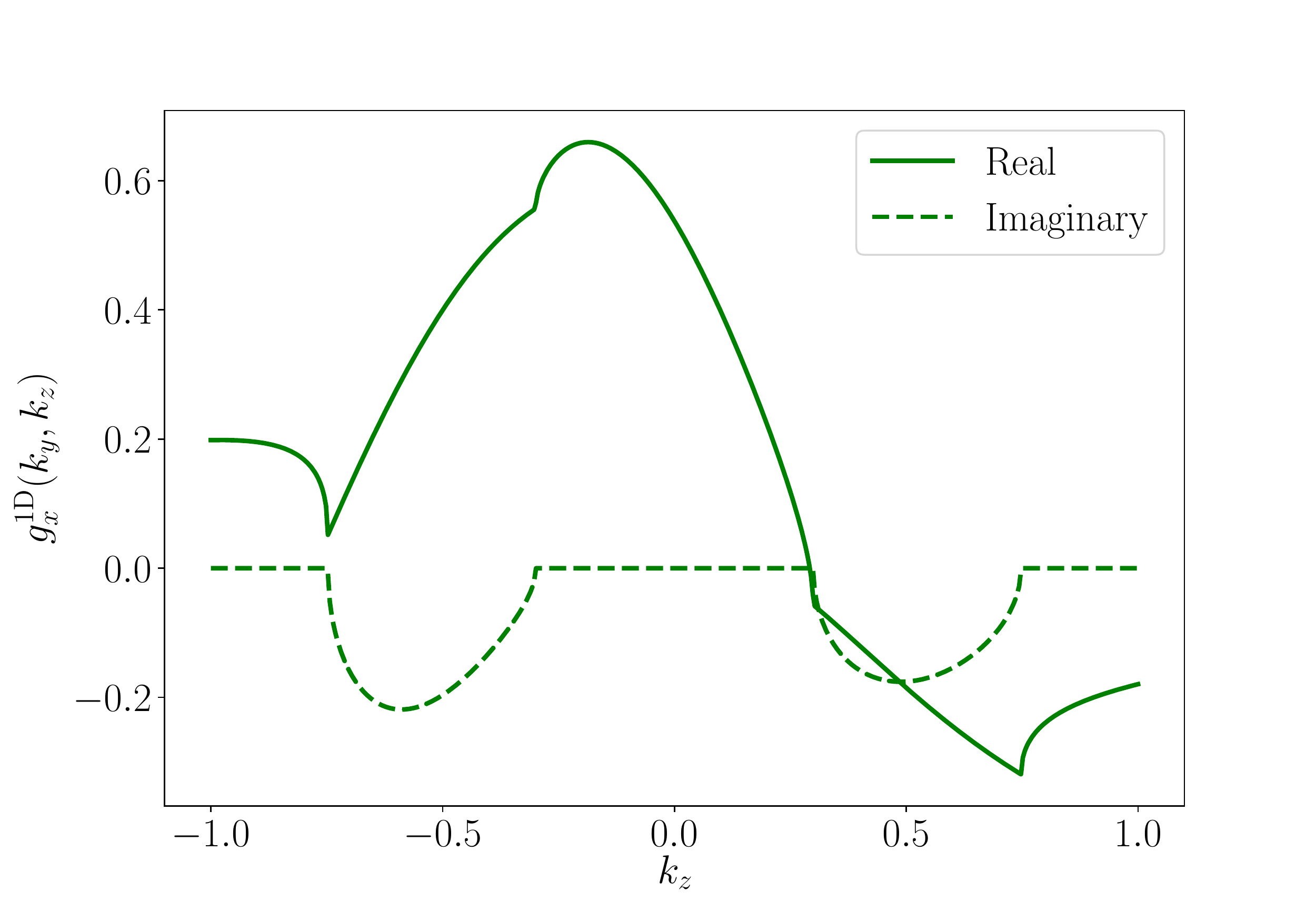}}
\caption{Typical integrand in Eq.~\eqref{eq:bulk_gf_ft3}. $[g^{\rm 1D}_x(k_y, k_z)]_{11}$ is plotted as a function of $k_z$ at $k_y = 0$ along the green dotted line in Fig.~\ref{fig:LDOS_Weyl}a for the surface of a Weyl semi-metal.}
\label{fig:1D_integrand}
\end{figure}

To integrate over $k_y$ and $k_z$ we use regular numerical integration (quadrature) where the interval $[-\pi,+\pi ]$ is broken up into small subintervals and the integrand is approximated by a polynomial in each subinterval.
A typical example of the integrand is shown in Fig.~\ref{fig:1D_integrand}. We observe that it possesses a number of cusps and kinks, so that integration over momentum is somewhat delicate.
To obtain reliable results in a robust way, we have found that the doubly adaptative scheme of Ref.~\onlinecite{quad_pedro} is particularly efficient.
Of prime importance is the algorithm used to evaluate the residual error of the integral, that allows one to compute $g^\mrm{3D}_{x,y,z}(E)$ to a given precision.
We plan to release an efficient implementation of the resulting algorithm as open source software\cite{C.Groth-in-preparation}.

\section{The glueing sequence} \label{sec:modified_systems}

The Dyson equation can be used to introduce modifications to the TIS Green's function.
This approach, that we call pictorially ``the glueing sequence'', has been discussed in different contexts~\cite{Bruno_RGF_2005, knit, molecular_electronics, fetter2012quantum}.
We follow the presentation of Ref.~\onlinecite{knit}.
An alternative, equivalent,  formalism uses the so-called T-matrix approach\cite{PhysRevB.69.134517, PhysRevB.86.115433}.

Let us suppose that the Hamiltonian of the system splits into two contributions $\hat H$ and $\hat W$.
Typically $\hat H$ is the Hamiltonian of the TIS system and $\hat W$ a perturbation that involves only a {\it finite}  number of sites.
(However, we shall see below that we can be slightly less restrictive.)
We suppose that we have already calculated the Green's function of the TIS,
\begin{equation}
  \hat g(E) = \frac{1}{E - \hat{H}  + i \eta},
\end{equation}
and we seek the Green's function of the full Hamiltonian,
\begin{equation}
\hat G(E) = \frac{1}{E - (\hat{H} + \hat{W}) + i \eta}.
\end{equation}
The starting point is a simple equation that is valid for any
two matrices $A$ and $B$:
\begin{equation}
\frac{1}{A+B} = \frac{1}{A} - \frac{1}{A} B \frac{1}{A+B}.
\end{equation}
In the present context, it translates into
\begin{equation}
\label{eq:dyson}
\hat G = \hat g  + \hat g \hat W \hat G,
\end{equation}
which is known as the Dyson equation. Eq.~\eqref{eq:dyson} involves infinite matrices. Let us introduce the set of points $\Omega$
that is the support of the matrix $\hat W$: $\forall i,j \notin \Omega, \ W_{ij}=0$. The crucial property of the Dyson equation is
that its projection on any finite set of points $\Theta$ that contains $\Omega$ forms a close set of equations,
\begin{equation}
\forall i,j\in\Theta, \ \ \ \hat G_{ij} = \hat g_{ij}  + \sum_{k,l\in\Omega} \hat g_{ik} \hat W_{kl} \hat G_{lj},
\end{equation}
that can be solved using standard linear algebra routines. More formally, we introduce the projectors $P_\Omega$ and $P_\Theta$ defined
as
\begin{equation}
P_\Theta = \sum_{i\in\Theta} \sum_{\mu \nu} |i, \mu\rangle\langle i, \nu|
\end{equation}
with a similar definition for $P_\Omega$. The finite matrix $\hat G_{\Omega\Theta} = P_\Omega \hat G P_\Theta$ is then given by
\begin{equation}
\hat G_{\Omega \Theta} = \frac{1}{1 -\hat  g_{\Omega\Omega} \hat W_{\Omega\Omega}}\hat  g_{\Omega \Theta}.
\end{equation}
In a second step, one calculates
\begin{equation}
  \hat G_{\Theta\Theta} = P_\Theta \hat G P_\Theta
=\hat  g_{\Theta\Theta} +\hat  g_{\Theta\Omega}\hat W_{\Omega\Omega}\hat G_{\Omega\Theta}.
\end{equation}
The above sequence allows to introduce any finite number of modifications to a system whose corresponding Green's function elements
are known.
This is a straightforward generalization of the basic sequence at the
core of the recursive Green's function algorithm.
Denoting be $N_\Omega$ and $N_\Theta$ the number of elements of the respective sets, the computational
cost of the glueing sequence scales as $N_\Omega N_\Theta^2$ (for non-local properties such as transport) or $N_\Omega^2 N_\Theta$ (for local properties such as local density of states).
This cost is independent of the system size (which is infinite) and only depends on the number of modifications and points of interest.

In this article, we use the glueing sequence in the following situations.

{\it Point-like impurities. } The simplest application of the glueing sequence is to modify selected on-site energies or hopping elements.
In this case, one strictly follows the sequence given above.
An example of such a calculation is the inclusion of on-site disorder in the application to Weyl semi-metals discussed in Sec.~\ref{sec:Weyl}.
If one is interested in calculating averages over
impurity configurations, then only the last stage of the calculation, the glueing sequence, needs to be recalculated for each sample.
Hence, these averages can be relatively cheap computationally.
A similar approach has been used previously in the context of disordered
graphene\cite{Ostrovsky_2010,Schelter_2011}.
In Ref.~\onlinecite{Ostrovsky_2010,Schelter_2011} the TIS Green's function has been obtained analytically through a route that is similar to the route that we have followed numerically.

{\it Attaching two systems together } The glueing sequence can also be used to connect two systems together, e.g. the semi-infinite 2D graphene sheet and the graphene nanoribbon shown in Fig.~\ref{fig:2D_leads}.
In this case the Dyson equation has an additional structure,
as the unperturbed Greeen's function is block-diagonal (the two subsystems are initially unconnected).
This structure may be used to simplify the algebra of the glueing sequence, see Ref.~\onlinecite{knit}.

{\it Creating a multilayer system.} The matrix $\hat W_2$ introduced in Sec.~\ref{sec:MTIS} has an infinite number of terms,
but it conserves momentum along the two directions $y$ and $z$.
It can be cast in the form
\begin{equation}
\hat W_2 = \int_{-\pi}^{\pi} \frac{dk_ydk_z}{(2\pi)^2} \sum_{i,j \in \Omega }W_{ij} |k_y,k_z,i\rangle\langle k_y,k_z,j|.
\end{equation}
Hence, for a given value of $k_y,k_z$, i.e.\ {\it before} we have performed the numerical integration on these variables, $\hat W_2$
takes the form of a finite perturbation and we may apply the glueing sequence. Such a step allows to create, e.g., multilayer systems.
The above equation is actually not fully general. A straightforward extension is to consider matrices $W_{ij}(k_y,k_z)$ that include
an explicit dependance on momentum. Terms of the class $\hat W_1$ are treated in a similar fashion.

{\it Slicing a system in two.} A particular example of perturbation of the $\hat W_2$ class is when the perturbation is exactly opposite to one
bond $V_x$.
\begin{equation}
\hat W_2 =  \sum_{y,z,\mu,\nu}  -[V_x]_{\mu\nu}|0,y,z,\mu\rangle\langle 1,y,z,\nu|+ h.c.
\end{equation}
In that case the perturbation slices the system and creates two disconnected systems. This is what we use to create the semi-infinite 2D graphene sheet as well as the surface of the Weyl semi-metal. It is important to notice that upon slicing, one may create bound states (1D), edge states (2D) or surface states (3D). In particular in topological materials, these states (that we refer collectively as bound states) are always present. Bound states may also appear in the multilayer example above. Properly dealing with bound states is the topic of the next section.
In the applications shown in this article, we focus on surfaces in the [100] direction.
The same technique can be extended to other surfaces, such as [110], by simply enlarging the unit cell so that [110] becomes effectively [100] for the large unit cell.

\section{The bound state problem} \label{sec:bound_state}
The last problem that remains to be addressed is the appearance of bound states in the system. These states appear in a wide variety of situations
that include slicing the system (creation of a surface), multilayers (quantum wells), Josephson junctions or around impurities.
In the present article, the edge states of the quantum spin Hall effect and the surface states (Fermi arcs) of the Weyl semi-metals are (generalized) bound states.  Bound states do not hybridize with the continuum and must therefore be handled as separate contributions. In MTIS, bound states may be invariant by translation upon zero (true bound states), one (edge states) or two directions (surface states). To simplify the discussion, we focus below on surface states. The results can be straightforwardly extended to the other situations.

\begin{figure}
\centerline{\includegraphics[width=93mm]{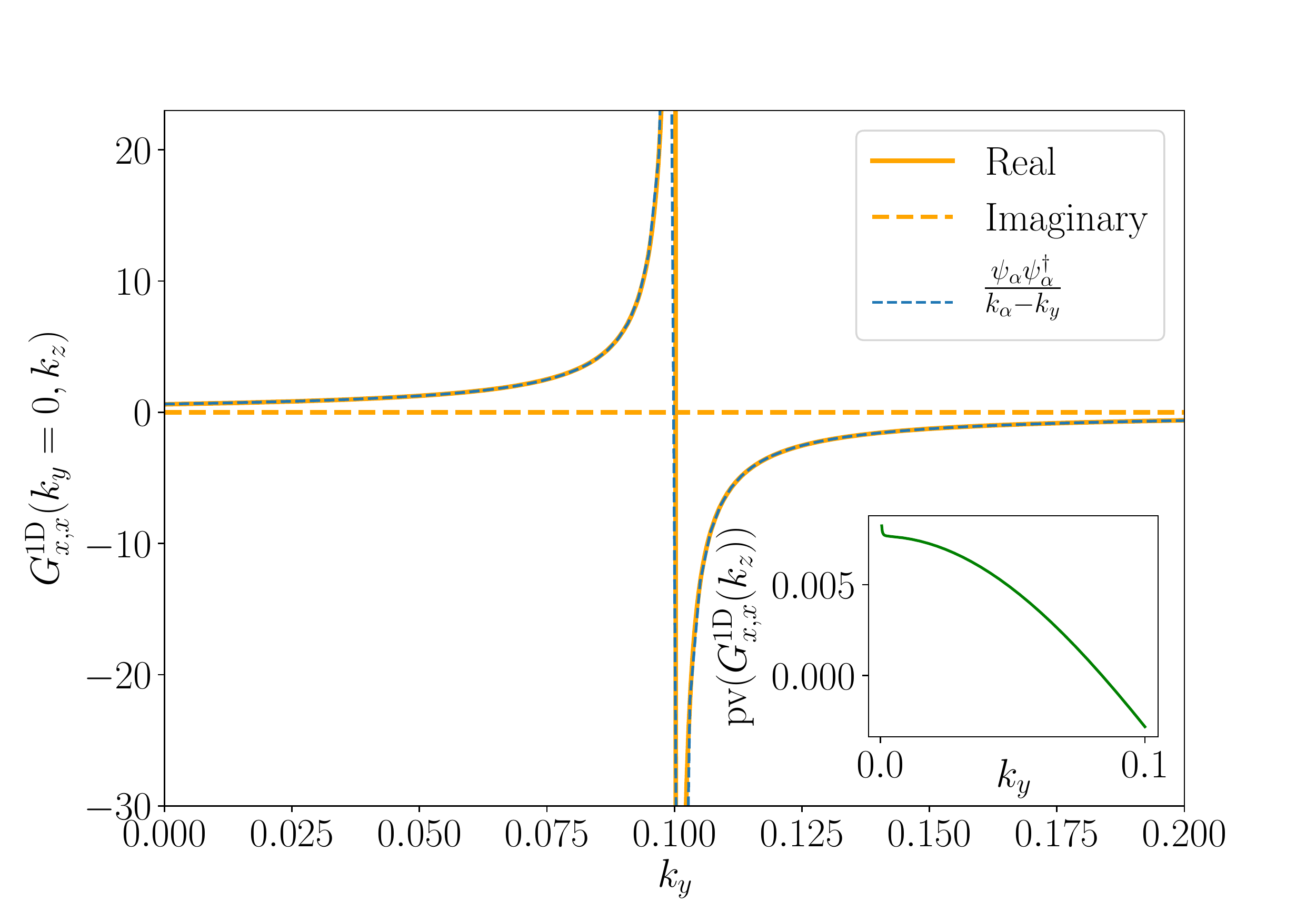}}
\caption{Behavior of the integrand around the surface state (Fermi arc) of the Weyl semi metal. $[G^{\rm 1D}_x(k_y, k_z)]_{22}$ as a function of $k_y$ at $k_z = 0$ along the orange dotted line in Fig.~\ref{fig:LDOS_Weyl}a. The fit $\psi_\alpha \psi_\alpha^\dagger / (\overline{k}_y - k_y)$ is only indicative and is not used in the analysis. The inset shows the regularized integrand, see text.}
\label{fig:1D_principal_value}
\end{figure}

Suppose that we have used the residue solver to construct the TIS Green's function $g_x(E,k_y,k_z)$ for fixed values of the transverse momentum
$(k_y, k_z)$. In a second step, we have used the glueing sequence to slice the system and obtain $G_{x,x'}(E,k_y,k_z)$, where the system terminates at $x=0$
(Fig.~\ref{fig:workflow}c). The contribution of a bound state $\psi_\alpha (k_y,k_z)$ with energy $E(k_y,k_z)$ to $G_{x,x'}(E,k_y,k_z)$ takes the form
\begin{equation}
\lim_{\eta \to 0} \frac{\psi_{\alpha}(x, k_y,k_z) \psi_{\alpha}^\dagger(x', k_y,k_z)}{E - E_\alpha(k_y,k_z) + i\eta}.
\end{equation}
Since $\lim_{\eta \to 0} 1/(X+i\eta)$ equals $P(1/X) -i\pi \delta(X)$,
the presence of a bound state leads to two complications: (i) The principal value
cannot be integrated with numerical routines as  the integral is formally divergent.
(ii) The Dirac function associated with the divergence at $E = E_\alpha(k_y,k_z)$ is not captured by $G_{x,x'}(E,k_y,k_z)$. Numerically, one only observes a numerical instability of
$G_{x,x'}(E,k_y,k_z)$ for values of $(k_y,k_z)$ such that $E = E_\alpha(k_y,k_z)$. An example of the integrand that results after the slicing sequence is
shown in Fig.~\ref{fig:1D_principal_value} where we indeed observe the behaviour discussed above.

To proceed, one needs first to calculate the bound states independently. We have devoted an earlier article to this problem to which we refer for all technical details\cite{Bound_state_algo}. We suppose that we have computed the energy $E_\alpha(k_y,k_z)$ and the associated state $\psi_{\alpha}(x, k_y,k_z)$ of the semi-infinite problem. Once this is done, we numerically inverse the function $E_\alpha(k_y,k_z)$ to obtain $k_\alpha = \bar k_y(E,k_z)$. Our goal is to perform the integration over $k_y$ and calculate
\begin{equation}
G_{xy,x'y'}(E,k_z) = \int_{-\pi}^{+\pi} \frac{dk_y}{2\pi} e^{ik_y (y-y')} G_{x,x'}(E,k_y,k_z).
\end{equation}

{\it (i) Regularizing the principal value.} The problem of the presence of the principal value is easily handled once one has evaluated the function
$k_\alpha = \bar k_y(E,k_z)$. We follow the standard procedure for handling principal values and select a small interval $[k_\alpha - \epsilon, k_\alpha - \epsilon]$. In this interval, we use the symmetrized integrand
\begin{equation}
\int_{k_\alpha - \epsilon}^{k_\alpha + \epsilon} \frac{dk_y}{2\pi} F(k_y)=
\int_{0}^{+ \epsilon} \frac{dq}{2\pi} [F(k_\alpha+q) +  F(k_\alpha-q)],
\end{equation}
where $F(k_y)=e^{ik_y (y-y')} G_{x,x'}(E,k_y,k_z)$.
The symmetrized integrand that gives the principal value of the real part plotted in Fig.~\ref{fig:1D_principal_value} is displayed in the inset of the same figure.
We find that although the original integral was divergent, this procedure correctly regularizes the integral.
Substracting $\psi_\alpha \psi_\alpha^\dagger / (k_\alpha - k_y)$ from $G_{x,x'}(E,k_y,k_z)$ could also regularize the integral.
Indeed, the dashed line of Fig.~\ref{fig:1D_principal_value} perfectly fits the divergence of the Green's function at the bound state.

{\it (ii) Contribution from the Dirac function.} We now evaluate the contribution of the bound state due to the Dirac function,
\begin{equation}
  \int_{-\pi}^{+\pi} \frac{dk_y}{2i} e^{ik_y (y-y')} \psi_{\alpha}(x, k_y) \psi_{\alpha}^\dagger(x', k_y)\delta[E - E_\alpha(k_y)],
\end{equation}
where we have droped the dependance on $k_z$ for compactness.
Performing this integral, we arrive at
\begin{equation}
\frac{1}{2i} e^{i k_\alpha (y-y')} \psi_{\alpha}(x, k_\alpha) \psi_{\alpha}^\dagger(x', k_\alpha)
\left|\frac{\partial E_\alpha}{\partial k_y}\right|^{-1}.
\end{equation}
The last step is to calculate $\partial E_\alpha/\partial k_y$.
In the case of a simple slicing where the system is invariant by translation away from the boundary, it is given by
\begin{align}
\frac{\partial E_\alpha}{\partial k_y}
& = \psi_\alpha^\dagger \frac{\hat H_{\rm 1D}}{\partial k_y} \psi_\alpha \\
& = \frac{i}{1- |\lambda_\alpha|^2} \psi_\alpha^\dagger(x=0) [V_y^\dagger e^{ik_y} - V_y e^{-ik_y}]\psi_\alpha(x=0),\nonumber
\end{align}
where $\lambda_\alpha$ is the evanescent momentum of the bound state.
For more complicated cases, e.g.\ a coated surface, there is a contribution from the part that is not invariant by translation, that plays the role of the scattering region in Ref.~\onlinecite{Bound_state_algo}.

\begin{widetext}
To summarize this section, the result of the integral over $k_y$ is given by
\begin{equation}
\begin{split}
G_{xy,x'y'}(E,k_z) = {} &
\frac{1}{2i} e^{i k_\alpha (y-y')} \psi_{\alpha}(x, k_\alpha) \psi_{\alpha}^\dagger(x', k_\alpha)
\left|\frac{\partial E_\alpha}{\partial k_y}\right|^{-1}
+\int_{k_y\notin [k_\alpha - \epsilon, k_\alpha - \epsilon]} \frac{dk_y}{2\pi} e^{ik_y (y-y')} G_{x,x'}(E,k_y,k_z) \\
& +
e^{ik_\alpha (y-y')} \int_{0}^{+ \epsilon} \frac{dq}{2\pi}
\left[ e^{iq (y-y')} G_{x,x'}(E,k_\alpha+q,k_z)  + e^{-iq (y-y')} G_{x,x'}(E,k_\alpha-q,k_z)  \right].
\end{split}
\end{equation}
Each of these integrals can now be performed using the numerical techniques discussed in Sec.~\ref{sec:periodic_2D_sys}.
\end{widetext}

\section{Conclusion} \label{sec:conclusion}
Despite being ubiquitous, quantum transport simulations face severe limitations in a number of situations where large systems are to be simulated.
This is in particular the case for most 3D systems.
The Mostly Translational Invariant Systems (MTIS) that we have discussed in this article cover a significant fraction of systems of interest that were unaccessible to simulations up to now.
We have presented a general method that can handle arbitrary MTIS, and demonstrated the power of the approach in a number of situations. The advantage of the MTIS approach stems from the fact that one works directly in the thermodynamic limit. Not only does the computational cost not depend on the size of the (infinite) system, but it is also independent of distances such as the distance between two impurities. It is therefore a natural tool for systems that possess several different characteristic length scales. The accuracy of the method is determined by that of the numerical integration and can be tuned at will.

We have studied the transport properties of the surface of a Weyl semi-metal.  In our calculation, there is a single surface present as the opposite second surface that would be present in a finite size calculation is sent infinitely far away.
Removing the second surface greatly simplifies the calculations and their interpretations. Our approach could be used for other surface problems such as the formation of Majorana bound states around magnetic impurities on top of a superconductor\cite{Cren_2015} or the study of the Dirac cones at the surface of a 3D topological insulators\cite{hasan_2010} or more generally for simulating
scanning tunneling microscope (STM) experiments.

For quantum transport, MTIS provides the possibility to use 2D or 3D electrodes such as the semi-infinite graphene sheet that we have presented. These electrodes are often more realistic than the quasi-1D electrodes used in almost all approaches. They are also less computationally intensive in most situations. We expect these electrodes to quickly become a standard feature of quantum transport toolkits.

MTIS also covers many applications that we have barely discussed so far. For pure bulk systems, they could be used for the study of defects, (RKKY) interaction between magnetic impurities or to calculate the collision integrals that enter the semi-classical Boltzman equation. The MTIS approach can be trivially combined with the recursive Green's function approach and its generalizations to build complex geometries.

\section*{Acknowledgments} 
Warm thank to Arthur Waintal for making Fig.\ref{fig:Weyl_moving_leads}a. We acknowledge interesting discussions with Anton Akhmerov and
Michael Wimmer.

This project has been funded by the US Office of Naval Research (ONR) and the graphene flagship through the ANR GRANSPORT.

\appendix

\section{Laurent expansion of the inverse of a matrix} \label{App:Laurent_exp}
Let $A(x)$ be a matrix that depends on a continuous parameter $x$. $A(x)$ is defined in terms of its expansion in power of $x$,
\begin{equation}
A(x) = A_0 + x A_1 + x^2 A_2 + \dots .
\end{equation}
Let $G$ be the inverse of $A$,
\begin{equation}
A(x)G(x) = \mathds{1}.
\label{eq:A_G}
\end{equation}
The aim of this appendix is to calculate the terms $G_k$ of the Laurent expansion of $G(x)$,
\begin{equation}
G(x) = \sum_{k = - \infty}^{\infty} G_k x^k,
\end{equation}
in terms of the expansion of $A(x)$. If $A_0$ is inversible, this problem is trivial: one can simply insert the expansions of $A(x)$ and $G(x)$ into Eq.~\eqref{eq:A_G} and identify the terms one by one.
One obtains
\begin{align}
G_0 & = A_0^{-1}, \\
G_1 & = -A_0^{-1} A_1 A_0^{-1}, \\
G_n & = -A_0^{-1}\sum_{p=0}^{n-1} A_{n-p} G_p.
\label{eq:expansion_of_A}
\end{align}

In this appendix, we focus on the case where $A_0$ is not invertible which leads to the appearance of a term $G_{-1}$. Let $P$ be the projector on the Kernel of $A_0$, i.e. $A_0P = 0$, and $Q = \mathds{1}-P$. To proceed, we write Eq.~\eqref{eq:A_G} in block form in the $P$
and $Q$ subblocks:
\begin{equation*}
(A_0 + x A_1 + x^2 A_2 + ...) (P + Q) G = \mathds{1}.
\end{equation*}
This expands to
\begin{equation*}
\left[ A_1 + x A_2 + ...) x P + (A_0 + x A_1 + x^2 A_2 + ...)Q \right]G = \mathds{1}.
\end{equation*}
Introducing the new variable
\begin{equation}
\overline{G} = (Px + Q) G,
\label{eq:G_bar}
\end{equation}
and the new series
\begin{equation}
B(x) \equiv B_0 + B_1 x + B_2 x^2 + ...
\end{equation}
with
\begin{equation}
B_k = A_{k+1} P + A_k Q.
\label{eq:B_definition}
\end{equation}
allows to map the problem of calculating $\overline{G}$ to a problem that
has the same structure as Eq.~(\ref{eq:A_G}):
\begin{equation}
B(x)\overline{G}(x) = \mathds{1}.
\label{eq:B_G}
\end{equation}
If $B_0 = A_1 P + A_0 Q = A_1 P + A_0$ is invertible, then a term-by-term identification of
Eq.~\eqref{eq:B_G} leads to
\begin{align}
\overline{G}_0 & = B_0^{-1}, \\
\overline{G}_1 & = -B_0^{-1} B_1 B_0^{-1}, \\
\overline{G}_n & = -B_0^{-1}\sum_{p=0}^{n-1} B_{n-p} \overline{G}_p. 
\end{align}
Eq.~\eqref{eq:G_bar} can be inverted to
\begin{equation}
G = \left(\frac{1}{x} P + Q \right) \overline{G}
\end{equation}
and we finally arrive at
\begin{align}
G_{-1} = {} & P\frac{1}{A_1 P + A_0} \label{eq:G-1},\\
G_0 = {} & -P\frac{1}{A_1 P + A_0} (A_2P+A_1Q)\frac{1}{A_1 P + A_0} \nonumber \\
 & + Q\frac{1}{A_1 P + A_0}, \\
G_n = {} & P \overline{G}_{n+1} + Q \overline{G}_{n}.
\label{eq:expansion_of_A2}
\end{align}
Eq.~\eqref{eq:G-1} is the central result used for the construction of the residue solver.
In all the calculations that we have performed, we have always found $B_0$ to be inversible.
If $B_0$ is not inversible, the same construction that has been applied to Eq.~\eqref{eq:A_G} can be performed on Eq.~\eqref{eq:B_G}. In that case, one would introduce the projector $P'$ onto the kernel of $B_0$ and proceed to expand $\overline{\overline{G}}=x P'\overline{G} + (1-P')
\overline{G}$. In that case, the developpement of $G(x)$ would start at $G_{-2}$. The same procedure can be extended iteratively to the case where the development starts at $G_{-n}$.

\bibliographystyle{unsrt}
\bibliography{Infinite_GF.bib}

\begin{thebibliography}{10}

\bibitem{PhysRevLett.47.882}
Patrick~A. Lee and Daniel~S. Fisher.
\newblock Anderson localization in two dimensions.
\newblock {\em Phys. Rev. Lett.}, 47:882--885, Sep 1981.

\bibitem{Thouless_1981}
D~J Thouless and S~Kirkpatrick.
\newblock Conductivity of the disordered linear chain.
\newblock {\em Journal of Physics C: Solid State Physics}, 14(3):235--245, jan
  1981.

\bibitem{MacKinnon1985}
A.~MacKinnon.
\newblock The calculation of transport properties and density of states of
  disordered solids.
\newblock {\em Zeitschrift f{\"u}r Physik B Condensed Matter}, 59(4):385--390,
  Dec 1985.

\bibitem{Bruno_RGF_2005}
G.~Metalidis and P.~Bruno.
\newblock Green's function technique for studying electron flow in
  two-dimensional mesoscopic samples.
\newblock {\em Phys. Rev. B}, 72:235304, Dec 2005.

\bibitem{Baranger_1991}
Harold~U. Baranger, David~P. DiVincenzo, Rodolfo~A. Jalabert, and A.~Douglas
  Stone.
\newblock Classical and quantum ballistic-transport anomalies in
  microjunctions.
\newblock {\em Phys. Rev. B}, 44:10637--10675, Nov 1991.

\bibitem{PhysRevB.44.8017}
T.~Ando.
\newblock Quantum point contacts in magnetic fields.
\newblock {\em Phys. Rev. B}, 44:8017--8027, Oct 1991.

\bibitem{kwant_article}
Christoph~W Groth, Michael Wimmer, Anton~R Akhmerov, and Xavier Waintal.
\newblock Kwant: a software package for quantum transport.
\newblock {\em New Journal of Physics}, 16(6):063065, 2014.

\bibitem{agrawal1993cutting}
Ajit Agrawal, Philip Klein, and R~Ravi.
\newblock Cutting down on fill using nested dissection: provably good
  elimination orderings.
\newblock In {\em Graph Theory and Sparse Matrix Computation}, pages 31--55.
  Springer, 1993.

\bibitem{Xu613}
Su-Yang Xu, Ilya Belopolski, Nasser Alidoust, Madhab Neupane, Guang Bian,
  Chenglong Zhang, Raman Sankar, Guoqing Chang, Zhujun Yuan, Chi-Cheng Lee,
  Shin-Ming Huang, Hao Zheng, Jie Ma, Daniel~S. Sanchez, BaoKai Wang, Arun
  Bansil, Fangcheng Chou, Pavel~P. Shibayev, Hsin Lin, Shuang Jia, and M.~Zahid
  Hasan.
\newblock Discovery of a weyl fermion semimetal and topological fermi arcs.
\newblock {\em Science}, 349(6248):613--617, 2015.

\bibitem{kpm_method}
Alexander Wei\ss{}e, Gerhard Wellein, Andreas Alvermann, and Holger Fehske.
\newblock The kernel polynomial method.
\newblock {\em Rev. Mod. Phys.}, 78:275--306, Mar 2006.

\bibitem{fan2018linear}
Zheyong Fan, Jose~Hugo Garcia, Aron~W Cummings, Jose-Eduardo Barrios, Michel
  Panhans, Ari Harju, Frank Ortmann, and Stephan Roche.
\newblock Linear scaling quantum transport methodologies.
\newblock {\em arXiv preprint arXiv:1811.07387}, 2018.

\bibitem{lanczos1950iteration}
Cornelius Lanczos.
\newblock {\em An iteration method for the solution of the eigenvalue problem
  of linear differential and integral operators}.
\newblock United States Governm. Press Office Los Angeles, CA, 1950.

\bibitem{low_T_Lanczos}
Markus Aichhorn, Maria Daghofer, Hans~Gerd Evertz, and Wolfgang von~der Linden.
\newblock Low-temperature lanczos method for strongly correlated systems.
\newblock {\em Phys. Rev. B}, 67:161103, Apr 2003.

\bibitem{finite_T_lanczos}
J.~Jaklič and P.~Prelovšek.
\newblock Finite-temperature properties of doped antiferromagnets.
\newblock {\em Advances in Physics}, 49(1):1--92, 2000.

\bibitem{Mayou_transport}
Pierre Darancet, Valerio Olevano, and Didier Mayou.
\newblock Quantum transport through resistive nanocontacts: Effective
  one-dimensional theory and conductance formulas for nonballistic leads.
\newblock {\em Phys. Rev. B}, 81:155422, Apr 2010.

\bibitem{PRL_Wingreen}
Yigal Meir and Ned~S. Wingreen.
\newblock Landauer formula for the current through an interacting electron
  region.
\newblock {\em Phys. Rev. Lett.}, 68:2512--2515, Apr 1992.

\bibitem{elke2017molecular}
Scheer Elke and Cuevas~Juan Carlos.
\newblock {\em Molecular electronics: an introduction to theory and
  experiment}, volume~15.
\newblock World Scientific, 2017.

\bibitem{datta_1995}
Supriyo Datta.
\newblock {\em Electronic Transport in Mesoscopic Systems}.
\newblock Cambridge Studies in Semiconductor Physics and Microelectronic
  Engineering. Cambridge University Press, 1995.

\bibitem{knit}
K.~Kazymyrenko and X.~Waintal.
\newblock Knitting algorithm for calculating green functions in quantum
  systems.
\newblock {\em Phys. Rev. B}, 77:115119, Mar 2008.

\bibitem{Bound_state_algo}
M.~Istas, C.~Groth, A.~R. Akhmerov, M.~Wimmer, and X.~Waintal.
\newblock A general algorithm for computing bound states in infinite
  tight-binding systems.
\newblock {\em SciPost Phys.}, 4:26, 2018.

\bibitem{crommie1993imaging}
MF~Crommie, CP~Lutz, and DM~Eigler.
\newblock Imaging standing waves in a two-dimensional electron gas.
\newblock {\em Nature}, 363(6429):524, 1993.

\bibitem{harrison1980solid}
Walter~A Harrison.
\newblock {\em Solid state theory}.
\newblock Courier Corporation, 1980.

\bibitem{Bernevig1757}
B.~Andrei Bernevig, Taylor~L. Hughes, and Shou-Cheng Zhang.
\newblock Quantum spin hall effect and topological phase transition in hgte
  quantum wells.
\newblock {\em Science}, 314(5806):1757--1761, 2006.

\bibitem{Quantum_transport_in_Weyl}
Weizhe~Edward Liu, Ewelina~M. Hankiewicz, and Dimitrie Culcer.
\newblock Quantum transport in weyl semimetal thin films in the presence of
  spin-orbit coupled impurities.
\newblock {\em Phys. Rev. B}, 96:045307, Jul 2017.

\bibitem{Weyl_t_matrix_impurity}
Andrew~K. Mitchell and Lars Fritz.
\newblock Signatures of weyl semimetals in quasiparticle interference.
\newblock {\em Phys. Rev. B}, 93:035137, Jan 2016.

\bibitem{Das_sarma_disorder_weyl}
Justin~H. Wilson, J.~H. Pixley, David~A. Huse, Gil Refael, and S.~Das~Sarma.
\newblock Do the surface fermi arcs in weyl semimetals survive disorder?
\newblock {\em Phys. Rev. B}, 97:235108, Jun 2018.

\bibitem{Fabry_perot_graphene}
Pierre Darancet, Valerio Olevano, and Didier Mayou.
\newblock Coherent electronic transport through graphene constrictions:
  Subwavelength regime and optical analogy.
\newblock {\em Phys. Rev. Lett.}, 102:136803, Mar 2009.

\bibitem{Vazifeh2013Electromagnetic}
M.~M. Vazifeh and M.~Franz.
\newblock Electromagnetic response of weyl semimetals.
\newblock {\em Phys. Rev. Lett.}, 111:027201, Jul 2013.

\bibitem{Buttiker1995}
M.~B{\"u}ttiker.
\newblock Time-dependent current partition in mesoscopic conductors.
\newblock {\em Il Nuovo Cimento B (1971-1996)}, 110(5):509--522, May 1995.

\bibitem{Wimmer_thesis}
Michael Wimmer.
\newblock Quantum transport in nanostructures: From computational concepts to
  spintronics in graphene and magnetic tunnel junctions, December 2009.

\bibitem{xu2015discovery}
Su-Yang Xu, Nasser Alidoust, Ilya Belopolski, Zhujun Yuan, Guang Bian, Tay-Rong
  Chang, Hao Zheng, Vladimir~N Strocov, Daniel~S Sanchez, Guoqing Chang, et~al.
\newblock Discovery of a weyl fermion state with fermi arcs in niobium
  arsenide.
\newblock {\em Nature Physics}, 11(9):748, 2015.

\bibitem{deng2016experimental}
Ke~Deng, Guoliang Wan, Peng Deng, Kenan Zhang, Shijie Ding, Eryin Wang, Mingzhe
  Yan, Huaqing Huang, Hongyun Zhang, Zhilin Xu, et~al.
\newblock Experimental observation of topological fermi arcs in type-ii weyl
  semimetal mote 2.
\newblock {\em Nature Physics}, 12(12):1105, 2016.

\bibitem{PhysRevB.93.235127}
E.~V. Gorbar, V.~A. Miransky, I.~A. Shovkovy, and P.~O. Sukhachov.
\newblock Origin of dissipative fermi arc transport in weyl semimetals.
\newblock {\em Phys. Rev. B}, 93:235127, Jun 2016.

\bibitem{PhysRevLett.113.026602}
Bj\"orn Sbierski, Gregor Pohl, Emil~J. Bergholtz, and Piet~W. Brouwer.
\newblock Quantum transport of disordered weyl semimetals at the nodal point.
\newblock {\em Phys. Rev. Lett.}, 113:026602, Jul 2014.

\bibitem{annurev-conmatphys-033117-054037}
Sergey~V. Syzranov and Leo Radzihovsky.
\newblock High-dimensional disorder-driven phenomena in weyl semimetals,
  semiconductors, and related systems.
\newblock {\em Annual Review of Condensed Matter Physics}, 9(1):35--58, 2018.

\bibitem{PhysRevLett.115.076601}
J.~H. Pixley, Pallab Goswami, and S.~Das~Sarma.
\newblock Anderson localization and the quantum phase diagram of three
  dimensional disordered dirac semimetals.
\newblock {\em Phys. Rev. Lett.}, 115:076601, Aug 2015.

\bibitem{PhysRevLett.115.246603}
Chui-Zhen Chen, Juntao Song, Hua Jiang, Qing-feng Sun, Ziqiang Wang, and X.~C.
  Xie.
\newblock Disorder and metal-insulator transitions in weyl semimetals.
\newblock {\em Phys. Rev. Lett.}, 115:246603, Dec 2015.

\bibitem{quad_pedro}
Pedro Gonnet.
\newblock {\em Adaptive quadrature re-revisited}.
\newblock PhD thesis, ETH Zurich, 2009.

\bibitem{C.Groth-in-preparation}
Groth C.
\newblock {\em In preparation.}

\bibitem{molecular_electronics}
Juan~Carlos Cuevas and Elke Scheer.
\newblock {\em Molecular Electronics}.
\newblock WORLD SCIENTIFIC, 2010.

\bibitem{fetter2012quantum}
Alexander~L Fetter and John~Dirk Walecka.
\newblock {\em Quantum theory of many-particle systems}.
\newblock Courier Corporation, 2012.

\bibitem{PhysRevB.69.134517}
Cristina Bena, Sudip Chakravarty, Jiangping Hu, and Chetan Nayak.
\newblock Quasiparticle scattering and local density of states in the
  d-density-wave phase.
\newblock {\em Phys. Rev. B}, 69:134517, Apr 2004.

\bibitem{PhysRevB.86.115433}
Annica~M. Black-Schaffer and Alexander~V. Balatsky.
\newblock Subsurface impurities and vacancies in a three-dimensional
  topological insulator.
\newblock {\em Phys. Rev. B}, 86:115433, Sep 2012.

\bibitem{Ostrovsky_2010}
P.~M. Ostrovsky, M.~Titov, S.~Bera, I.~V. Gornyi, and A.~D. Mirlin.
\newblock Diffusion and criticality in undoped graphene with resonant
  scatterers.
\newblock {\em Phys. Rev. Lett.}, 105:266803, Dec 2010.

\bibitem{Schelter_2011}
J.~Schelter, P.~M. Ostrovsky, I.~V. Gornyi, B.~Trauzettel, and M.~Titov.
\newblock Color-dependent conductance of graphene with adatoms.
\newblock {\em Phys. Rev. Lett.}, 106:166806, Apr 2011.

\bibitem{Cren_2015}
Gerbold~C M{\'e}nard, S{\'e}bastien Guissart, Christophe Brun, St{\'e}phane
  Pons, Vasily~S Stolyarov, Fran{\c{c}}ois Debontridder, Matthieu~V Leclerc,
  Etienne Janod, Laurent Cario, Dimitri Roditchev, et~al.
\newblock Coherent long-range magnetic bound states in a superconductor.
\newblock {\em Nature Physics}, 11(12):1013, 2015.

\bibitem{hasan_2010}
M.~Z. Hasan and C.~L. Kane.
\newblock Colloquium: Topological insulators.
\newblock {\em Rev. Mod. Phys.}, 82:3045--3067, Nov 2010.

\end{thebibliography}

\end{document}